\documentclass[aps,prd,preprint,showpacs,preprintnumbers,amsmath,amssymb,floatfix]{revtex4-1}
\usepackage{graphicx}
\usepackage{slashed}
\usepackage{subcaption}
\usepackage{amsmath}
\usepackage{amsfonts}
\usepackage{amssymb}
\usepackage{url}
\usepackage{amsbsy}
\usepackage{mathtools}

\usepackage[colorlinks]{hyperref}
\usepackage[usenames,dvipsnames]{color}

\usepackage[applemac]{inputenc}
\usepackage[T1]{fontenc}

\def\beq{\begin{equation}}
\def\eeq{\end{equation}}
\def\bea{\begin{eqnarray}}
\def\eea{\end{eqnarray}}
\def\be{\begin{equation}}
\def\ee{\end{equation}}
\def\ba{\begin{array}}
\def\ea{\end{array}}

\def\part{\partial}

\def\Tr{\mbox{Tr}}

\def\thup{\theta_\uparrow}
\def\thdn{\theta_\downarrow}
\def\sthup{\sinh\theta_\uparrow}
\def\cthup{\cosh\theta_\uparrow}
\def\sthdn{\sinh\theta_\downarrow}
\def\cthdn{\cosh\theta_\downarrow}

\def\evar{\varepsilon}
\def\vecp{\mathbf{p}}
\def\vecq{\mathbf{q}}

\def\vze{{\bf s}}
\def\ze{s}
\def\vecS{\vec{\mathcal{S}}}
\def\psl{\rlap{/}p}
\def\mbm{\mathbf{M}}

\def\be{\begin{equation}}
\def\ee{\end{equation}}
\def\bea{\begin{eqnarray}}
\def\eea{\end{eqnarray}}
\newcommand{\eV}{\text{eV}}

\def\sss{\scriptscriptstyle}

\def\pF{p_{\sss F}}

\begin{document}

\preprint{UdeM-GPP-TH-19-268}

\title{Cosmic Ferromagnetism of Magninos}
\author{R. B. MacKenzie$^{1}$}
\email{richard.mackenzie@umontreal.ca}
\author{M. B. Paranjape$^1$}
\email{paranj@lps.umontreal.ca}
\author{U. A. Yajnik$^{2}$}
\email{yajnik@iitb.ac.in}

\affiliation{$^1$Groupe de physique des particules, D\'epartement de
physique, Universit\'e de Montr\'eal, C.P. 6128, Succ. Centre-ville,
Montr\'eal, Qu\'ebec, CANADA, H3C 3J7 }
\affiliation{$^2$ Physics Department, Indian Institute of Technology Bombay, Mumbai, 400076, India}

\begin{abstract}
We study the physical conditions for the occurrence of ferromagnetic instability in a neutral plasma of fermions. 
We consider a system of two species $M$ and $Y$ which are oppositely charged under a local $U(1)_{X}$, with $M$ much
lighter than $Y$. The leading correction to free quasiparticle behaviour for the lighter species arises from the 
exchange interaction, while the heavier species remain spectators. This plasma, which is abelian, asymmetric and idealised, 
is shown to be naturally susceptible to the formation of a completely spin-imbalanced 
ferromagnetic state for the lighter species (dubbed a magnino) in large parts of parameter space. It is shown that
the domain structure formed by this ferromagnetic state can mimic Dark Energy, determining the masses of the
two fermion species involved, depending on their abundance relative to the standard photons. Incomplete cancellation 
of the X-magnetic fields among the domains can give rise to residual long range $X$-magnetic fields.
Under the assumption that this $U(1)_{X}$ mixes with Maxwell electromagnetism, this provides a mechanism 
for the seed for cosmic-scale magnetic fields. An extended model with several flavours $M_a$ and $Y_a$ of the species 
can incorporate Dark Matter. Thus the scenario shows the potential for explaining the large scale magnetic fields, and 
what are arguably the two most important  outstanding puzzles of cosmology: Dark Matter and Dark Energy.
\end{abstract}

\maketitle

\tableofcontents

\section{Introduction}
There are several important unresolved issues in our current understanding of cosmology. Paramount among these are the problems 
of Dark Matter (DM) and Dark Energy (DE). Of these, DM assists in galaxy formation and it seems consistent
for it to be a gas of non-relativistic particles throughout the epoch of galaxy formation, although the nature  
of such particles and their interaction with standard matter are not yet understood.
On the other hand, the issue of DE is closely tied to that of the cosmological constant \cite{1989-Weinberg-Rev.Mod.Phys.} 
and could be a hint of a new fundamental constant of physics.  
Indeed, the $\Lambda$-CDM model that best fits the cosmic microwave background (CMB) data \cite{2018-Aghanim.others-b}
suggests that it is a constant over the epochs scanned by the CMB. 
But assuming it is a dynamical phenomenon, it has to be assigned the equation of state $p=-\rho$ which demands an explanation 
in terms of relativistic  phenomena. From the point of view of naturalness, explaining a value of a dynamically generated quantity 
which is many orders of magnitude 
away from any of the scales of elementary particle physics or gravity is a major challenge. 

A class of possible explanations rely  on  scenarios involving extra 
dimensions or stringy physics \cite{2011-Li.etal-Commun.Theor.Phys.}. 
A more conventional explanation for DE may be expected to arise from  some  species of 
particles  that admits a nontrivial ground state which in the long-wavelength regime simulates  
non-zero vacuum energy.   There exist several proposals along these lines which rely on dynamical symmetry breaking or 
principles known from low-temperature physics, for instance, suggesting a new phenomenon \cite{2004-Kapusta-Phys.Rev.Lett.} 
or an explanation of  DE \cite{2018-Dey.etal-Nucl.Phys.}.  An explanation of DE relying entirely on autonomous physics of 
some particle species will  also demand particles at such a low mass scale, which has become phenomenologically  justified 
since the neutrino sector has shown  the existence of a very low mass scale.  There also exist natural mechanisms to connect  
such a low scale to known high-scale physics, although these are yet to be verified. However due to several alternatives involved, 
one may first carry out an investigation agnostic of such  high scale connection.

In this paper
we pursue one such approach. We consider a new sector of particles with interaction mediated
by an unbroken abelian gauge symmetry denoted $U(1)_{X}$. The core of our mechanism involves the existence of
a fermionic species that  enters into a ferromagnetic state. As we will show, it is required to have an extremely small mass
and hence an extremely large magnetic moment; we dub this species the \textsl{magnino}, denoted $M$. We assume 
that the medium remains neutral under the $X$-charge due to the presence of a significantly heavier 
species $Y$ of opposite charge which does not enter the 
collective ferromagnetic state. The existence of two such oppositely-charged species that do not mutually annihilate
is very much borrowed from known physics, and indeed below we will review previous studies of ferromagnetism in 
relativistic electron systems, and then will adapt them to our model. If we also assume additional flavours of the two types of 
particles and suitable flavour symmetries, it is possible to explain DM within the same sector,
including possible dark atoms formed by such species \cite{2009-Feng.etal-JCAP}\cite{2016-Boddy.etal-Phys.Rev.}
\cite{2014-Cline.etal-Phys.Rev.b,2012-Cline.etal-Phys.Rev.}.
This would also solve the \textit{concordance problem}, that is, the comparable energy densities carried in the cosmological
energy budget by the otherwise-unrelated components, DM and DE.

A linkage of the sector proposed here to the observed sector may exist through kinetic mixing of the $X$-electromagnetism with 
the standard one. The existence of cosmic magnetic fields at galactic and intergalactic scales 
\cite{Kulsrud:2007an}\cite{Durrer:2013pga}\cite{Subramanian:2015lua} is an outstanding puzzle of cosmology. 
Our mechanism relying as it does on spontaneous formation of domains of $X$-ferromgnetism has the potential to
provide the seeds needed to generate the observed fields through such mixing. 

Another possibility resulting from the mixing of the two $U(1)$'s is for the new particles to carry minicharges 
 \cite{1986-Holdom-Phys.Lett.,1986-Holdom-Phys.Lett.a} (originally dubbed millicharged).
Recently, it has been suggested that the excessive absorption of CMB in the era of early star formation
reported by EDGES \cite{2018-Bowman.etal-Nature} can be explained by the interaction of DM with standard matter
 \cite{2018-Barkana-Nature}, specifically of  minicharged DM, 
 \cite{2018-Munoz.etal-Phys.Rev.Lett.,2018-Munoz.Loeb-Nature}. Our scenario includes the possibility of minicharged 
particles through kinetic mixing, and could be compatible with this interpretation of the EDGES observation.
In \cite{2017-Daido.etal-Phys.Lett.} a proposal to
 embed minicharged particles within a unified theory is proposed. A proposal to search for minicharged particles 
for masses $<10^{-4}$eV and kinetic mixing of hidden photons with standard photons $\xi\sim10^{-8}$ 
has been made \cite{2014-Li.Voloshin-Mod.Phys.Lett.}, which is roughly the range of parameters natural to our scenario. 

The light particles needed by our magnino model could have arisen from the decay of long-lived heavier particles 
after Big Bang Nucleosynthesis (BBN). If however their existence preceded nucleosynthesis, there will be additional 
effective relativistic degrees of freedom at that time.  It has been suggested in \cite{2018-Poulin.etal-} that the existence of additional 
degrees of freedom may help resolve the discrepancy between
direct measurements of $H_0$ from the Hubble Space Telescope \cite{Freedman:2000cf} and Large scale structure 
\cite{2014-Betoule.others-Astron.Astrophys.}\cite{2018-Abbott.others-,2018-Abbott.others-a} , and high redshift "High-$z$" 
Type Ia supernova projects \cite{2018-Riess.others-Astrophys.J.} on the 
one hand and the CMB 
determination of $\Lambda$-CDM parameters by WMAP \cite{Komatsu:2010fb}\cite{Komatsu:2014ioa} and Planck \cite{2018-Aghanim.others-b} satellite experiments on the other. 

From a variety of experiments, MiniBoone\cite{Aguilar-Arevalo:2018gpe}, IceCube \cite{2018-Aartsen.others-Phys.Rev.Lett.}
and other experiments have suggested the existence of sterile neutrinos, that is, non-standard 
light fermionic species with interactions outside the Standard Model (SM). It as been pointed out that the MiniBoone results demand an
explanation beyond mere mixing with SM neutrinos\cite{Aguilar-Arevalo:2017mqx}\cite{2018-Liao.etal-, 2018-Jordan.etal-}. 
At least some of these could be accommodated within the proposal made here. We comment on this possibility in the concluding section.

Thus, we offer an utterly conventional solution to several of the puzzling issues in cosmology by relying on only  known phenomena in many body physics 
and replication of several features of the observed spectrum of elementary particles. More generally, the mechanism of ferromagnetism considered 
is shown to be operative at extremely low number densities and temperature close to absolute zero as obtains in our very late universe, 
provided a few of the intrinsic mass scales are minuscule. It would be interesting to explore this phase of matter in its own right,
as possibly applicable to a variety of fermionic species of observed as well as potentially of the "hidden" sector. 
Specifically, relying as it does on a strongly correlated system but coupled through a weak abelian gauge force, this model allows easy 
understanding and ready deployment of known strategies of verification for such a phenomenon.

It has been brought to our notice that Raby and West have introduced the term \textit{magnino} for a proposed Dark Matter particle interacting
by the Electroweak force, that could also simultaneously solve the solar neutrino puzzle \cite{1988-Raby.West-Phys.Lett.a}\cite{1987-Raby.West-Phys.Lett.}. Our particle  primarily enters  into the solution of the Dark Energy puzzle, and we justify the usage of the suffix \textit{-ino} to mean an 
extremely light fermion. We proceed to
use this term in our context for convenience, and suggest that of the two distinct proposed species
the one first to receive confirmation of existence may acquire this term permanently. 
We begin this paper in Sec.~\ref{sec:cosmo} with a recapitulation of the current status of cosmology and the way 
negative pressure  may be seen to arise in Sec.~\ref{sec:negpress}. Sec.~\ref{sec:rfg} reviews the many-body theory result for ferromagnetic 
instability of a relativistic electron gas. In Sec.~\ref{sec:dw} we discuss the phenomenology of domain wall (DW) formation 
and evolution in a cosmological setting. In Sec \ref{sec:singlemagnino} we describe the main proposal of this paper, adapting the 
calculation presented earlier to the case of interest, where rather than electrons and nuclei (the latter being essentially bystanders 
to keep the system neutral) we have in mind magninos $M$ and their oppositely charged heavy cousins $Y$. This system forms a plasma which is 
asymmetric, abelian and idealised (referred to as a PAAI). We identify the parameter range for which a collective phase of magninos 
and the attendant fate of the $Y$ particles can together be identified with DE.  In Sec.~\ref{sec:flavo} we discuss augmented versions 
of this sector in which additional heavier fermions can act as
DM and help to resolve the concordance puzzle. In Sec.~\ref{sec:cosmagfields} we discuss the
possibility that the same DE scenario could provide seeds for the intergalactic magnetic fields in 
the Universe. In Sec.~\ref{sec:conclusion} we summarize our results and discuss future avenues of research.

\section{Cosmological setting}
\label{sec:cosmo}
Determining the nature of the DE component of matter in the universe observed through distant candles 
\cite{Riess:1998cb,Perlmutter:1998np}  and inferred from observations such as the
CMB  precision data WMAP \cite{Komatsu:2010fb,Komatsu:2014ioa} and Planck \cite{Ade:2015xua}\cite{2018-Aghanim.others-b} 
presents a new challenge to fundamental 
physics. The exceedingly small mass scale associated with this energy density
makes it unnatural to interpret it as a cosmological constant \cite{1989-Weinberg-Rev.Mod.Phys.}, and therefore
demands an unusual mechanism for relating it to the physics of known
elementary particles. On the other hand, a new window to very-low-mass physics 
has opened up with the discovery of the low mass scale of neutrinos. (For reviews,
see \cite{Bahcall:2004ut}, \cite{Maltoni:2004ei}\cite{Bergstrom:2015rba}).
Furthermore, a variety of theoretically motivated ultra-light species are currently
being sought experimentally \cite{Jaeckel:2010ni,2010-Jaeckel.Ringwald-Ann.Rev.Nucl.Part.Sci.}. We may therefore exploit the
presence of an ultra-light sector to explain the DE phenomenon autonomously 
at a low scale without direct reference to its high-scale connection with known physics.

\subsection{Sources of negative pressure}
\label{sec:negpress}
\begin{figure}[thb]
\begin{subfigure}{.4\textwidth}
  \includegraphics[width=0.9\linewidth]{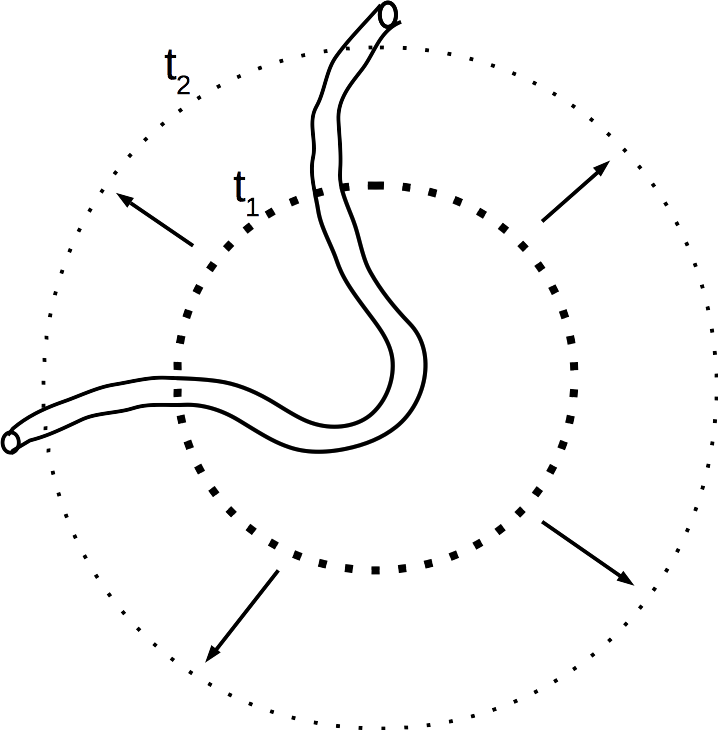}
  \label{fig:scalingvortex}
\caption{}
\end{subfigure}
\begin{subfigure}{.4\textwidth}
  \includegraphics[width=0.9\linewidth]{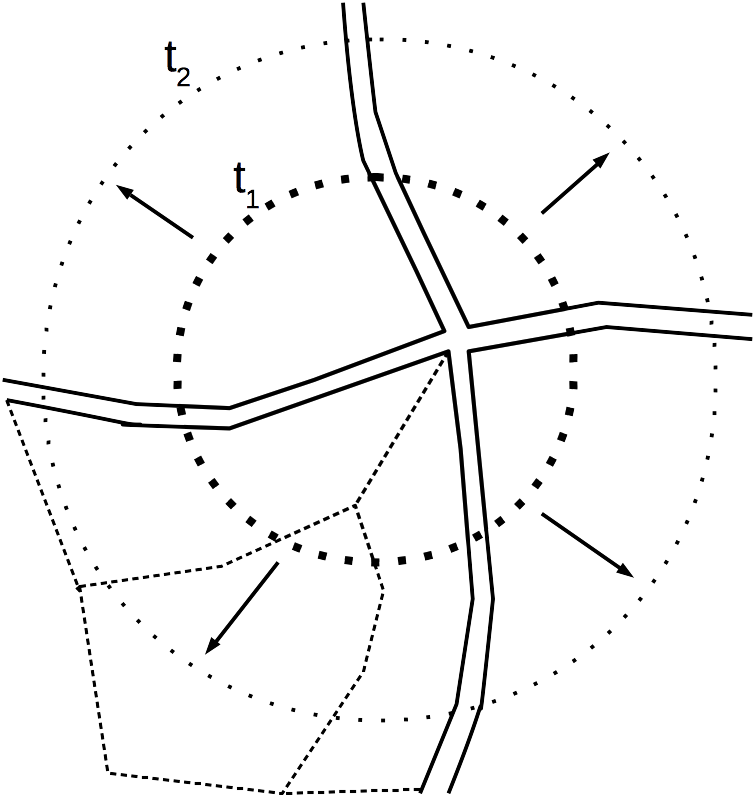}
  \caption{}
  \label{fig:scalingdomainwall}
\end{subfigure}
\caption{Scale factor increasing during time $t_1$ to $t_2$ around a slow-moving vortex in the left panel and
around a wall junction in the right panel.  The wall junction is schematically shown in cross-section, with dashed lines depicting wall areas recessed from the page. 
The volume demarcated by the scale factor at a later time $t_2$  contains vortex length 
proportional to the scale factor $a$, while in the right panel the wall area contained in this volume is proportional to $a^2$.
These factors partially offset the volume dependence of material density $\propto a^{-3}$, resulting
in energy density $\propto 1/a^2$ in vortex network and $\propto 1/a$ in domain wall complex.}
\label{fig:scalingvortexandwall}
\end{figure}

\begin{figure}[bht]
  \includegraphics[width=.6\linewidth]{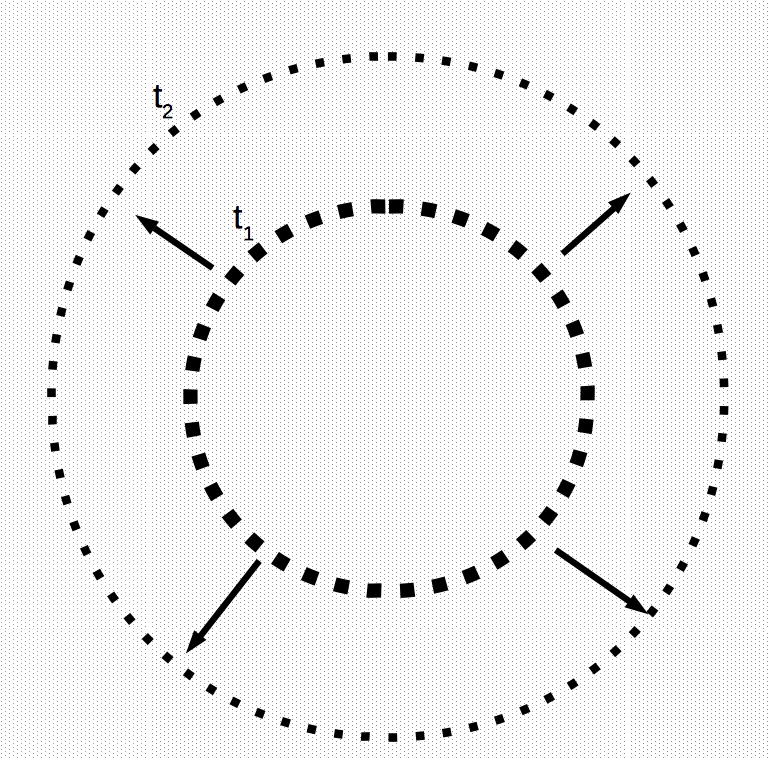}
  \caption{Scale factor increasing during time $t_1$ to $t_2$ in a space filled with a relativistic scalar. The volume demarcated by the scale factor at later value of time $t_2$ contains additional amount of the scalar substance proportional to the volume $a^3$. This factor completely offsets the volume dependence of material density $\propto a^{-3}$, providing constant energy density at all times despite expanding universe.}
  \label{fig:scalingvacuum}
\end{figure}

A homogeneous, isotropic universe at critical density  is described by the Friedmann equation for the scale factor $a(t)$
\beq
\left( \frac{1}{a}\frac{da}{dt} \right)^2 = \frac{8\pi}{3} G \rho
\eeq
and the covariant conservation of energy-momentum, 
\beq
 \frac{d}{da}\left(\rho a^3 \right) +3 p a^2= 0.
\label{eq:covcons}
\eeq
This needs to be supplemented by an equation of state relation $p=w\rho$ where
$w$ is constant for a universe dominated by a given type of matter. Alternatively, if one has auxiliary knowledge 
of $\rho[a]$ as a functional of the geometric scale factor $a$ then this equation determines the functional 
$p[a]$. The simplest known sources are the relativistic gas with $w=1/3$ and the non-relativistic gas, with $w=0$. However the existence
of extended relativistic objects in gauge theories allows for novel possibilities. In Fig.s~\ref{fig:scalingvortexandwall} 
we have sketched the growth of the scale factor in the presence of different extended objects.
Vortices and domain walls usually form an extended network or a complex as first discussed
by Kibble\cite{Kibble:1980mv}.  After an initial transient phase of their evolution, they become slow-moving with
relatively low kinetic energy per unit length or area, especially true for domain walls as they must
be attached to each other forming a wall complex. In this case their kinetic contribution to pressure is 
negligible just as in the case of non-relativistic particles. However, the energy density scales as
$1/a^3$ for nonrelativistic particles, while this scaling changes for extended objects. The cartoons in Fig.~\ref{fig:scalingvortexandwall} suggest how the effective energy density to be included in the above calculations is to be deduced. In the case of a frozen-out vortex line network, the situation is quite similar to non-relativistic particles, increasing their average separation as $1/a^3$. 
But as sketched in the figure, there is also an increment in the energy proportional to $a$ due to an average 
length of vortex network proportional to $a$ entering the physical volume. As such, the energy density of the 
network has to be taken to scale as $1/a^2$. Likewise, for a domain wall complex, there is a monotonous 
increase of energy proportional to $a^2$, making the effective energy density proportional to $1/a$. Finally, 
if we have a relativistically meaningful space-filling extended substance which is homogenous, 
the expansion in the scale factor causes energy proportional to $a^3$ to be included, as shown in 
Fig.~\ref{fig:scalingvacuum}. This makes its energy density contribution remain
constant as $a$ grows. In quantum theory this arises naturally as the vacuum expectation
value of a relativistic scalar field.

If we now take the three scaling laws of the preceding paragraph and plug them into \eqref{eq:covcons}, we
can determine the corresponding pressure as $p[a]$, resulting in the following equations of state which are well-known in the cosmology literature \cite{KolTur,Dodelson}:
\beq
\begin{array}{ll}
p=-\frac{1}{3}\rho &\qquad \mathrm{network\  of\  vortices} \\
p=-\frac{2}{3}\rho  &\qquad \mathrm{complex\  of\  domain\  walls} \\
p=-\rho &\qquad \mathrm{homogeneous\  vacuum\  energy}
\end{array}
\eeq
In the following, we consider a scenario that gives rise to a complex of domain walls. The scale of such structure would
be set by the intrinsic dynamics determining the phase transition in which it arises. By comparison, the  causal horizon during the 
epochs in which the presence of Dark Energy  can be detected ( since the surface of last scattering) sets a very large scale, 
$10^{17}-10^{25}$metre. Thus  the  DW network is expected to exist on scales minuscule compared 
to cosmological length scales. We shall argue that subsequent to its initial formation, such structure remains frozen, retaining
constant physical size  till the later epochs. Hence our structure, with justifiable averaging, is more appropriately
represented by Fig.~\ref{fig:scalingvacuum}. This will make the rate at which the energy of new walls gets included in the 
fiducial volume grow as $a^3$. Hence our structure may justifiably be assumed to simulate an equation of state $p=-\rho$.


\subsection{Summary of cosmological data}
\label{sec:cosmosummary}
The present universe is well described by three components: non-relativistic matter which 
includes both Standard Model (SM) matter such as baryons and electrons and DM, relativistic
or quasi-relativistic SM species such as photons and neutrinos, and the DE. 
With these, the Friedmann equation becomes
\be
\label{eq:scalefactor}
\left( \frac{1}{a}\frac{da}{dt} \right)^2 = \frac{8\pi}{3} G 
\left( \rho_{\text{rel,}0}\left(\frac{a_0}{a}\right)^4 + \rho_{\text{m,}0}\left(\frac{a_0}{a}\right)^3 + \rho_\Lambda \right),
\ee
where the subscript $0$ refers to current values while subscripts rel, m and $\Lambda$ refer to relativistic matter, 
nonrelativistic matter and the cosmological constant, respectively.  The current Hubble constant is usually stated as \cite{PDG} 
$H_0=h\times 100 \text{ (km/s)/Mpc}$ with $h$ a parameter which according to direct observations such as that of the Hubble 
Space Telescope and Type Ia supernova is $0.72$ while according to CMB data of WMAP and Planck it is $0.678$.
For our purposes a precise value is unnecessary and we set $h^2=0.5$. 
The other parameters are the total energy density, the so-called critical value $\rho_\text{c}$ which agrees with 
this Hubble value and a spatially flat Universe, and  $n_\gamma$, the number density of the CMB photons.
Useful conversion factors and the parameter values we use are as follows:
\bea
\hbar c=1&=& 197 \text{ MeV F} \\ \nonumber
(\text{cm})^{-3}&=&7.6\times10^{-15}(\eV)^3 \\ \nonumber
h^2\equiv (H_0/(100\ \text{(km/s)/Mpc}))^2 &=&0.5 \qquad (\text{chosen for simplicity})\\ \nonumber
\rho_\text{c}=\left( \frac{3}{8\pi}G\right)(H_0)^2&=& 4.01\times10^{-11} (\eV)^4 \\ \nonumber
n_\gamma = 410\times\left(\frac{T_\gamma}{2.7255}\right)^3 (\text{cm})^{-3} &=& 3.12\times 10^{-12} (\eV)^3 \\ \nonumber
\rho_\text{DE}=0.70 \rho_\text{c}&=& 2.81\times10^{-11} (\eV)^4  \\ \nonumber
\rho_\text{DM}=0.26 \rho_\text{c}&=& 1.04\times10^{-11} (\eV)^4 
\eea
 
The analysis of the CMB experiments \cite{Komatsu:2014ioa}\cite{2018-Aghanim.others-b} assumes the $\Lambda$-CDM model, with  equation 
of state of the DE constrained to $p/\rho\equiv w=-1$. However, alternative 
analyses (see for instance \cite{Shafieloo:2009ti}) show that a dynamically evolving $w$ 
is also consistent with data. 
It has been argued, early in \cite{Battye:1999eq} that the equation of state 
obeyed by the observed contribution to the energy density could be well fitted by a 
network of frustrated domain walls \cite{Kibble:1980mv}, which obey an effective 
equation of state $p=(-2/3)\rho$ in the static limit.  
This  possibility was further examined 
in \cite{Conversi:2004pi} \cite{Friedland:2002qs} although it may not be  consistent  with more recent data. 
We mention these possibilities here more as examples of divergence from the consensus about
the nature of the DE. Our mechanism is compatible with $w=-1$ for the Dark Energy sector over at least
substantial part of the existence of that special phase. But it arising at a specific epoch within the observable
past  and also possibly degrading within recent visible epochs are possibilities we comment on in the conclusion section. 

Finally, as far as DM is concerned, from large-scale structure (LSS) data and also fit to the CMB 
data, it is known that the DM candidate particle must be non-relativistic by the time galactic structure formation starts.
This requires $M_{DM}$ to at least be in the keV range. It has been argued in \cite{Boyarsky:2008ju, Boyarsky:2008xj}, 
that the Lyman-alpha forest data corresponding to the early galaxies suggests that the mass of the DM candidate
could be as low as in the keV range. 
Our model provides a viable solution for all the mass values of the potential DM particles in the eV to GeV range and accords with this expectation.

\section{Ferromagnetic instability of relativistic Fermi gas}
\label{sec:rfg}
We begin with a  brief review of the various contexts in which a collective state such as the one 
being discussed here has been studied. 
In condensed matter physics, the more prevalent explanations of the ferromagnetic instability are 
based on exchange coupling between electrons in the orbitals of ions from neighbouring sites. This 
is the standard Heisenberg ferromagnetism arising from localised
electrons. By contrast the model we pursue is that of delocalised or band fermions.
This  is sometimes also referred to as the case of ``itinerant'' fermions.  In our scenario there is only gaseous indefinitely 
extensive medium. In the case of itinerant fermions in the background of a periodic lattice 
somewhat more interesting situation can arise. Such an ansatz for ferromagnetism was first proposed by Stoner \cite{Stoner372}. 
However the effective interaction proposed by this ansatz is not easy to derive from first principles \cite{Rajagopal:1973}. 

An early study of the collective states of a homogeneous neutral plasma including the 
interactions was done in Akhiezer and Peletminskii \cite{Akhiezer:1960} whose results have served as 
a benchmark for many subsequent calculations. Accordingly, in the formalism due to Baym and Chin 
\cite{Baym:1975va,Chin:1977iz} it is sufficient to perform a summation of the 
ladder diagrams arising from forward scattering due to the gauge field interaction.
A variety of collective phenomena have been widely explored also in the case of QCD by
Tatsumi \cite{Tatsumi:1999ab,Tatsumi:2008}, and in \cite{Son:2007ny} and \cite{Pal:2009yf}. 
At sufficiently high chemical potential such as in the 
interior of a neutron star, colour superconductivity as well as chromoferromagnetism have been 
proposed. These proposals are specialised to the non-abelian $SU(3)$ interactions. We shall be 
considering the abelian case, leading to one vital difference in the exchange energy contribution, as will be explained.

\subsection{Plasma which is asymmetric, abelian and idealised (PAAI)}
\label{sec:PAAI}
A system of fermions (conventional electrons in this section but magninos below) interacting through an abelian gauge 
force can be treated as a gas  of weakly 
interacting quasiparticles under certain conditions. The most useful setting is that of an electron 
gas being the more active dynamical medium in the presence of oppositely charged much heavier ions 
or protons which are mostly spectators and serve to keep  the medium neutral. The  total energy of 
such a system can be treated as a functional of electron number 
density, according to the Hohenberg-Kohn theorem.  
In a relativistic setting, it becomes a functional of the covariant 4-current, and hence also of 
the electron spin density \cite{Rajagopal:1973}.

\begin{figure}[bht]
  \includegraphics[width=\linewidth]{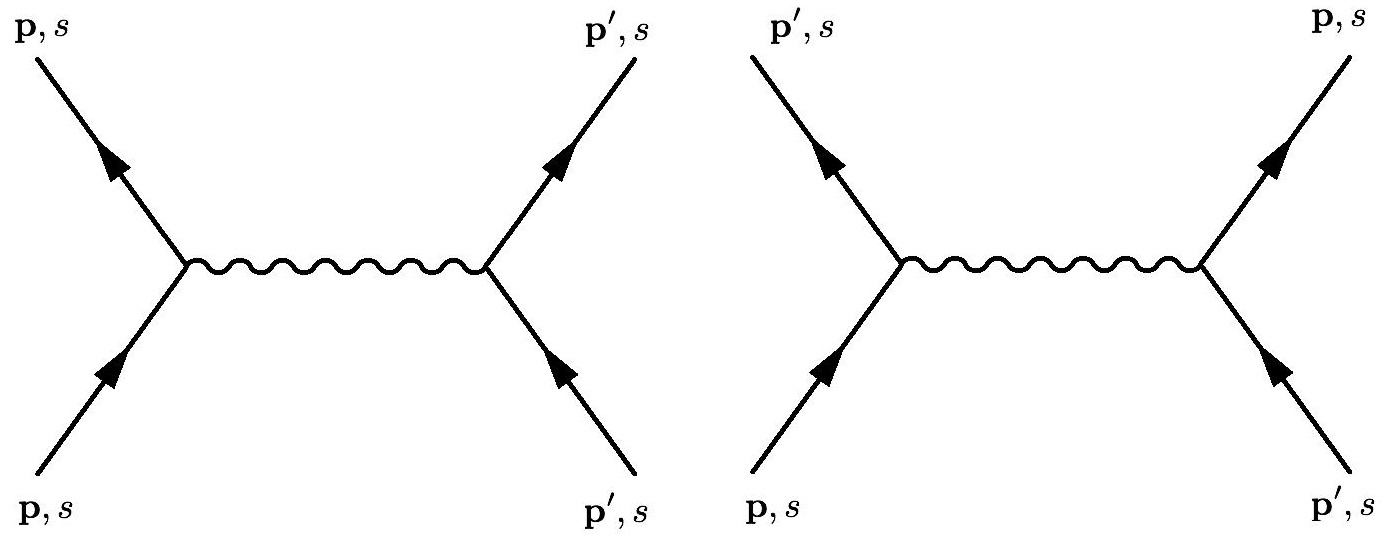}
  \caption{The two forward scattering diagrams contributing to exchange energy, in which the magnitudes of the momenta are the same.}
  \label{fig:fwdscattering}
\end{figure}

In the Landau liquid formalism further elucidated in \cite{Baym:1975va,Chin:1977iz},
the Coulomb interaction between the lighter fermions may be ignored due to shielding
provided by heavy oppositely charged partners which do not participate directly in dynamics. 
In our proposal this is the gas of the oppositely charged heavier particles. We discuss such a 
plasma in some detail, without the complications of lattice effects and accordingly call this
the \textit{ideal abelian asymmetric plasma} for convenience abbreviated PAAI where the qualifier asymmetry refers to 
the large ratio of the masses of the oppositely charged species.
When the Fermi liquid is considered for a spin polarised liquid, the total energy of the system $E$ is a 
functional of the phase space distribution function $n$ and spin $s$ of the quasiparticles, where the 
covariant convention for spin basis is discussed later. 
The quasiparticle energy spectrum $\evar$ in a volume $V$ is defined through
\beq
\delta E/V = \sum_{\ze=\pm} \int \evar(\vecp \ze) \delta n(\vecp \ze) d\tau \quad \text{where} \quad 
d\tau=d^3p/(2\pi \hbar)^3.
\eeq
The interaction strength $f(\vecp \ze, \vecp' \ze')$ between quasi-particles is defined 
as
\beq
\delta \evar(\vecp,\ze) = \sum_{\ze'=\pm}\int f(\vecp \ze, \vecp' \ze')\delta n(\vecp' \ze') d\tau'
\eeq
which is thus a second variation of the energy $E$, is symmetric in its 
arguments and vanishes in a non-interacting gas.
This phenomenological quantity can be related to the scattering
amplitude as follows. For electrons in a solid there are interactions mediated by  
residual electromagnetic interactions of a vector nature and scalar interactions mediated by phonons. 
Combined these contributions to scattering processes give rise to a
non-vanishing Green's Function $K(p_3,p_4,p_1,p_2)$ for the process 
$p_3,p_4, \leftarrow p_1,p_2$, where $p_i$ are four-vectors and spin labels are suppressed.  Its one-particle-irreducible piece $\Gamma(p_3,p_4, p_1, p_2)$ provides the required input to the effective theory of 
quasi-particle static quantities like the energy (see Sec.~15 of \cite{Lifshitz:1981stp}). Causality arguments can then be 
used to show that the function $f$  is given by the forward scattering limit of 
$\Gamma$:
\beq
f(\vecp \ze, \vecp' \ze') = \Gamma^{\omega}( \vecp \ze , \vecp' \ze' )
\label{eq:ftoGamma}
\eeq
where the function on the right is defined by 
\beq
\Gamma^{\omega}(\vecp \ze, \vecp' \ze') = \lim_{\vecq\rightarrow 0} 
\Gamma(\vecp+\mathbf{q} \ze, \vecp'-\mathbf{q} \ze', \vecp \ze, \vecp' \ze'), \qquad |\mathbf{q}|/\omega_q 
\rightarrow 0
\label{eq:Gammalimit}
\eeq 
The offshell forward scattering limit is prescribed to be taken in the unphysical domain with 
momentum $q$ going to zero before the corresponding energy is allowed to go to 
zero. 
In practice, \eqref{eq:ftoGamma} is put to use in the form 
 \cite{Baym:1975va}
\beq
f(\vecp\vze,\, 
\vecp'\vze')=\frac{m}{\evar^0(\vecp)}\frac{m}{\evar^0(\vecp')}\mathcal{M}(\vecp\ze,\, 
\vecp'\ze'),
\label{eq:fM}
\eeq
where $\evar^0$ is the free particle energy and $\mathcal{M}$ is the Lorentz-covariant $2\rightarrow2$ scattering amplitude in
the limit specified above. This is shown in Fig.~\ref{fig:fwdscattering}. The exchange energy can equivalently be seen to arise as a 
two-loop correction to the self-energy of the fermion \cite{Chin:1977iz}.

Next we discuss the spin basis following closely the presentation of \cite{Bu:1984}.  We shall assume that the thermal 
energy is much lower than the mass threshold of the lighter species
and that the spontaneous creation of anti-particles is suppressed. As such, the main relativistic effect seen is in the 
behaviour of the spin. Relativistic treatment of spin requires that  the 
spin basis functions are also momentum-dependent. 
We refer the spin to the rest frame of the fermion to begin with, where it 
is denoted $\vze$. The covariant projection operators needed to identify 
polarisation states are defined in terms of the vector boosted to the  
frame of the moving particle $\vecp$,
\beq
\mathcal{S}^0=\frac{\vecp\cdot\vze}{m}, \qquad 
\vecS=\vze+\frac{\vecp(\vecp\cdot\vze)}{m(E_p+m)};
\label{eq:Sdef}
\eeq
the projection operator for (positive energy) particles of spin 
$\vze$ is
\beq
\rho(p,s)=\frac{1}{2m}(\psl+m)\times \frac{1}{2}(1+\gamma^5 
\rlap{/}\mathcal{S})
\label{eq:densitymatrix}
\eeq
For noninteracting fermions, the Feynman propagator with nonzero particle 
density is \cite{Bu:1984}
\bea
S_F&=&\sum_{\pm\vze}\frac{({\psl}_+ +m)(1+\gamma^5 
\rlap{/}\mathcal{S}_+)}{4E_p}\left( 
\frac{n_{F\vze}(\vecp)}{p^0-E_p-i\eta}+\frac{1-n_{F\vze}(\vecp)}{p^0-E_p+i\eta}
\right)\\
&-&\sum_{\pm\vze}\frac{({\psl}_- +m)(1+\gamma^5 
\rlap{/}\mathcal{S}_-)}{4E_p}\left( 
\frac{1-\bar{n}_{F\vze}(\vecp)}{p^0+E_p-i\eta}+\frac{\bar{n}_{F\vze}(\vecp)}{
p^0+E_p+i\eta }
\right)T
\eea

This propagator can be used to recover the equilibrium density  $n_p$ as a function of 
momentum :
\bea
n_p&=&\int\frac{dp^0}{2\pi i}\Tr(\gamma^0S^{<}_F(p))\\
&=& \sum_{\pm \vze}\left[n_{F\vze}(\vecp)+(1-\bar{n}_F\vze(\vecp))  \right]
\eea
and the magnetisation vector $\mbm(\vecp)$,
\bea
\mbm(\vecp)&=&\mu_B\int \frac{dp^0}{2\pi i}\Tr(\gamma^0\vec{\Sigma} 
S^{<}_F(p))\\ \nonumber
&=&\mu_B\frac{m}{E_p}\sum_{\pm\vze}\int \frac{dp^0}{2\pi i} 
\left( 
\frac{\vec{\mathcal{S}}_+n_{F\vze}(\vecp)}{p^0-E_p-i\eta}-\frac{\vec{\mathcal{S}
} _-(1-\bar { n } _ { F\vze } (\vecp)) } { p^0-E_p+i\eta } \right)\\ \nonumber
&=&\mu_B\frac{m}{E_p}\sum_{\pm\vze}\left[\vec{\mathcal{S}}_+n_{F\vze}(\vecp) -  
\vec{\mathcal{S}}_-(1-\bar{n} _ { F\vze } (\vecp)) \right]\\ 
&=&\mu_B\frac{m}{E_p}\vec{\mathcal{S}}_+(n_{F\uparrow}(\vecp)-n_{F\downarrow}
(\vecp))
\label{eq:magnetisationformalism}
\eea
where $\mu_B$ is the Bohr magneton and in the last step we have set 
$\bar{n}_{F\vze}=0$.

To set up a spin-asymmetric state, we introduce a parameter $\zeta$ such that the 
net density $n$ splits up into densities of spin up and down fermions as
\beq
n_{\uparrow}=n(1+\zeta) \quad \mathrm{and}\quad n_{\downarrow}=n(1-\zeta) 
\eeq
Correspondingly, we have Fermi momenta $p_{F\uparrow}=p_F(1+\zeta)^{1/3}$ and $p_{F\downarrow}=p_F(1-\zeta)^{1/3}$, with 
$p_F^3=3\pi^2n$. In terms of these, the total magnetisation per unit volume 
can be found be
\beq
\mbm=\frac{\mu_Bm^3}{6\pi^2}{\vze} \left[ \frac{1}{3}\left(\frac{p_{\uparrow}}{m}\right)^3 
+ \frac{p_{\uparrow}}{m}\left(1+\left(\frac{p_{\uparrow}}{m}\right)^2\right)^{1/2}  
-\sinh^{-1}\left(\frac{p_{\uparrow}}{m}\right) - (p_{\downarrow}\leftrightarrow p_{\uparrow})
\right]
\label{eq:magnetisation}
\eeq
These quantities are calculated from the free-particle Green's function but with nonzero fermion number.

Next we turn to the calculation of the effective energy density of the fermion gas. There are two main 
contributions: (1) the kinetic energy of the quasiparticles with 
renormalised mass parameter and (2) the spin-dependent exchange energy of a spin-polarised gas. 
Contribution (2), denoted exchange energy $E_\text{xc}$, is
\beq
E_\text{xc} = \sum_{\pm \vze}\sum_{\pm \vze'}\int \frac{d^3 p}{(2\pi)^3}\frac{d^3 p'}{(2\pi)^3}
f(\vecp\vze,\, \vecp'\vze')n(\vecp,\vze)n(\vecp',\vze')
\eeq
Using \eqref{eq:fM} this can be expressed in the covariant form as the two terms
arising from the Feynman diagrams Fig. \ref{fig:fwdscattering} in the Coulomb gauge \cite{Bu:1984}
\beq
E_\text{xc}^\text{direct}(n,\alpha)=2\pi e^2\int\frac{d^4p}{(2\pi)^4} \int\frac{d^4p}{(2\pi)^4}
\frac{1}{|\vecp-\vecp'|^2}\Tr(\gamma^0S^{<}_F(p)\gamma^0S^{<}_F(p'))
\eeq
\beq
E_\text{xc}^\text{crossed}(n,\alpha)=2\pi e^2 \int\frac{d^4p}{(2\pi)^4} \int\frac{d^4p}{(2\pi)^4}
\frac{\delta_{ij}-q_iq_j/|\vecq|^2}{(p^0-p'^0)^2-|\vecq|^2}\Tr(\gamma^iS^{<}_F(p)\gamma^jS^{<}
_F(p'))
\eeq
where $\alpha=e^2/4\pi$ is the fine structure of ordinary matter and $\vecq=|\vecp-\vecp'|$. 
(Below we will replace $\alpha\to\alpha_{X}$, the corresponding fine structure constant of the hidden $U(1)_{X}$.) 
These quantities were calculated in \cite{Bu:1984} and the result is given next.
A similar calculation done for a quark liquid by Tatsumi \cite{Tatsumi:1999ab} does not have the term $E_\text{xc}^\text{direct}$ 
due to the vanishing of trace over colour degrees of freedom. Continuing, we compute the kinetic energy
and add to it the exchange energy.

For the spin-asymmetric state, we define the variables
\beq
\sinh \theta =\frac{p_F}{m}; \qquad \sinh \thup =\frac{p_{F\uparrow}}{m} \qquad \sinh \thdn =\frac{p_{F\downarrow}}{m}
\eeq
Then the kinetic energy is given by
\beq
E_\text{kin}=\frac{m^4}{16\pi^2}\left\{ \frac{1}{4}\sinh \thup - \thup + \frac{1}{4}\sinh \thdn - \thdn \right\} - \frac{m^4}{3 \pi^2} \sinh^3 \theta 
\label{eq:ekin}
\eeq
where the last term subtracted is the rest energy. In order to determine the preferred minimum of the collective state this
contribution is irrelevant, but we will need to restitute this term when dealing with gravity.
The exchange energy is given by \cite{Bu:1984}
\bea
E_\text{xc}&=&\frac{\alpha m^4}{2 \pi^2}\left\{ \frac{1}{4}(\sthup\cthup - \thup+\sthdn \cthdn -\thdn)^2 \right. \nonumber \\
&+& \frac{1}{3}(\sthup \cthup -\sthdn \cthdn + \thdn -\thup )(\thup-\thdn+\sinh \thdn-\sinh\thup)  \nonumber\\
&-& \frac{4}{3} (\sthup\cthup -\sthdn\cthdn)(\thup-\thdn)   \nonumber \\
&+&  \frac{7}{6} (\thup-\thdn)^2 + \frac{1}{2}(\sthup-\sthdn)^2   + \frac{1}{3} (\sthup-\sthdn)(\thup-\thdn) \nonumber \\
&+ &2(\sthup\sthdn +\thup\thdn - \thdn\sthup\cthup - \thup\sthdn\cthdn)  \nonumber \\
&-&\frac{2}{3}(\cthup-\cthdn)^2 \ln\left| \frac{\sinh((\thup-\thdn)/2)}{\sinh((\thup+\thdn)/2)}  \right|  \nonumber \\
&-&\left. \frac{1}{6}(\cthup+1)^2I(\thdn;\thup)-\frac{1}{6}(\cthdn+1)^2I(\thup;\thdn) \right\}
\label{eq:exchange}
\eea
where 
\beq
I(\theta_n;\theta_m) = \int_0^{\theta_n} du \frac{\sinh(u/2)}{\cosh(u/2)} \ln\left| \frac{\sinh((\theta_m-u)/2)}{\sinh((\theta_m+u)/2}  \right|
\eeq
for which one can prove the property that
\beq
I(\theta_n;\theta_m)+I(\theta_m;\theta_n)=-\theta_n\theta_m.
\eeq
We define $\beta=p_{_F}/m$ and $E=E_\text{kin}+E_\text{xc}$. The question now is whether the energy surface in the $\alpha$-$\beta$ 
plane (allowing $\alpha$ to vary in anticipation of applying the above calculation to the hidden sector)
prefers a ground state with $\zeta\neq0$. It is found that there is a competition between the kinetic energy which does not involve $\alpha$ 
and the $E_\text{xc}$ which can be monotonically negative as a function of $\zeta$. Thus given any $\beta$, there exists a value of $\alpha$
that will make $\zeta=1$ lower in energy than the point $\zeta=0$. Also, for any $\alpha$, we can make $\beta$ large enough that 
$E_\text{kin}$ dominates and  $\zeta=0$ remains the unique ground state.
 
\begin{figure}[t]
  \includegraphics[width=.8\linewidth]{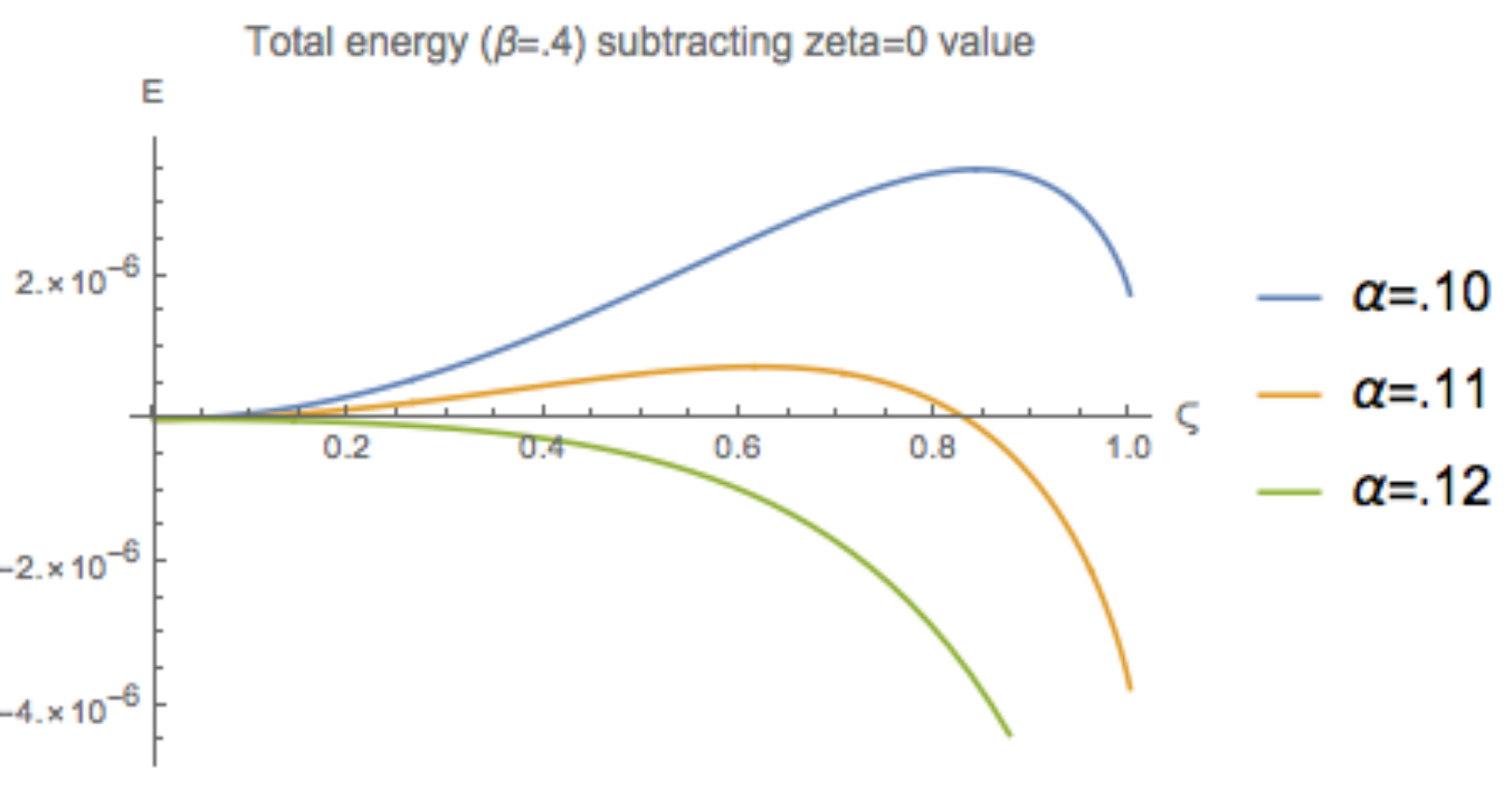}
  \caption{Energy as a function of the spin asymmetry parameter $\zeta$ near one of the critical 
values 
  of fine structure constant $\alpha$ with $\beta=p_F/m=0.4$.
  }
  \label{fig:alphavarying}
\end{figure}

\begin{figure}[t]
  \includegraphics[width=.8\linewidth]{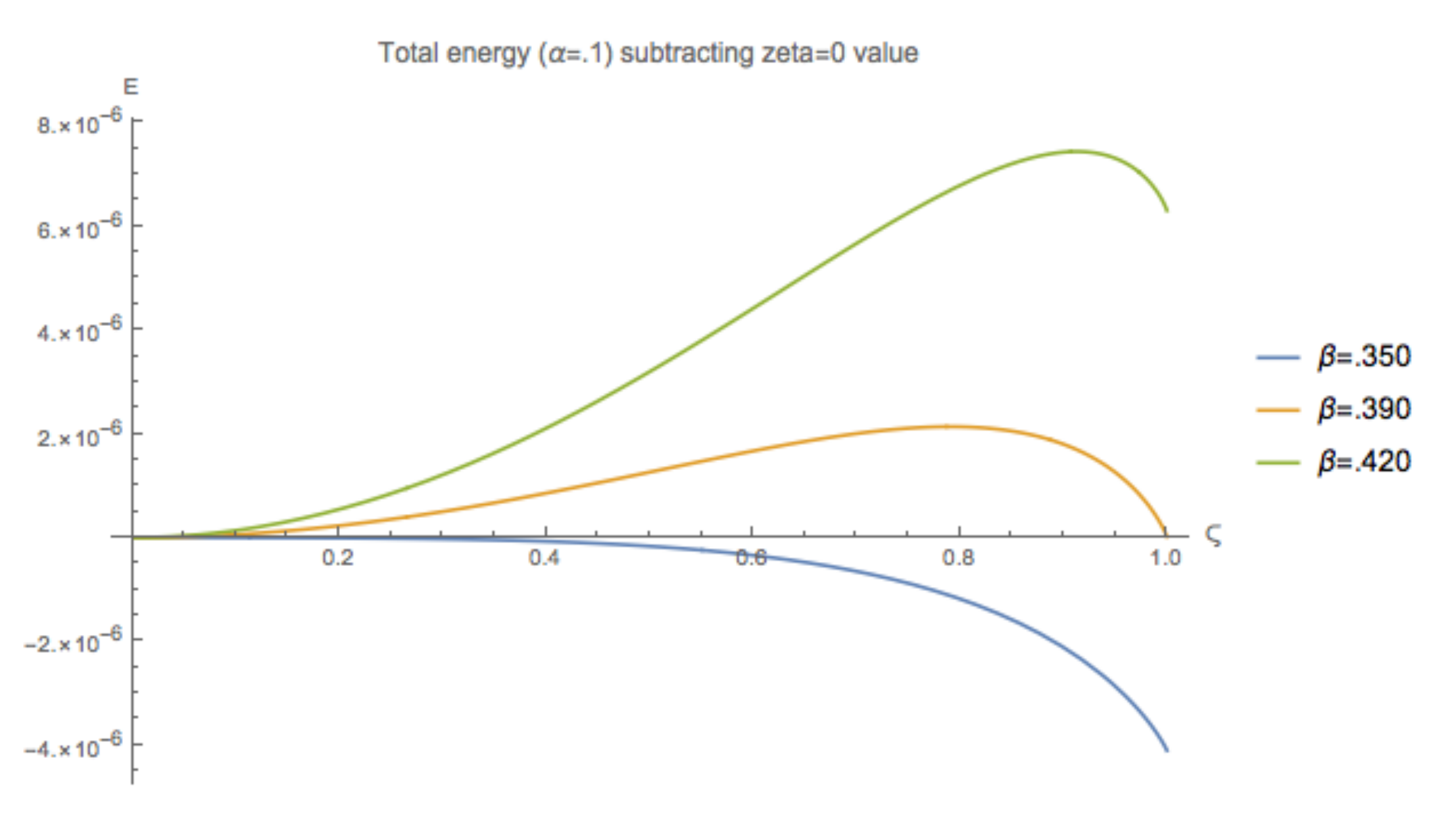}
  \caption{Energy as a function of the spin asymmetry parameter $\zeta$ near one of the critical 
values 
  of $\beta=p_F/m$ with fine structure constant $\alpha=0.1$.}
  \label{fig:betavarying}
\end{figure}

\subsection{Phase diagram and equation of state}
\label{sec:phasediaeos}
In Fig.~\ref{fig:alphavarying} we show an example of the
effect of increasing $\alpha$ with a fixed value of $\beta=0.4$. For convenience we have plotted 
$E(\zeta=1)-E(\zeta=0)$ to ensure the same origin on the
ordinate. $\alpha=0.10$ already shows turning over of the curve near $\zeta=1$, but that local minimum is metastable. For
$\alpha=0.11$, $\zeta=1$ is the lower minimum, with a barrier separating it from the local minimum at $\zeta=0$. But at $\alpha=0.12$,
the graph is monotonically concave down and $\zeta=1$ is the only minimum.
Likewise, for fixed $\alpha=0.1$, Fig.~\ref{fig:betavarying} shows the effect of varying $\beta$. The global minimum at 
$\zeta=0$ gets destabilised with decreasing $\beta$, passing through a small range of $\beta$ with two local minima. 

\begin{figure}[bt]
  \includegraphics[width=.8\linewidth]{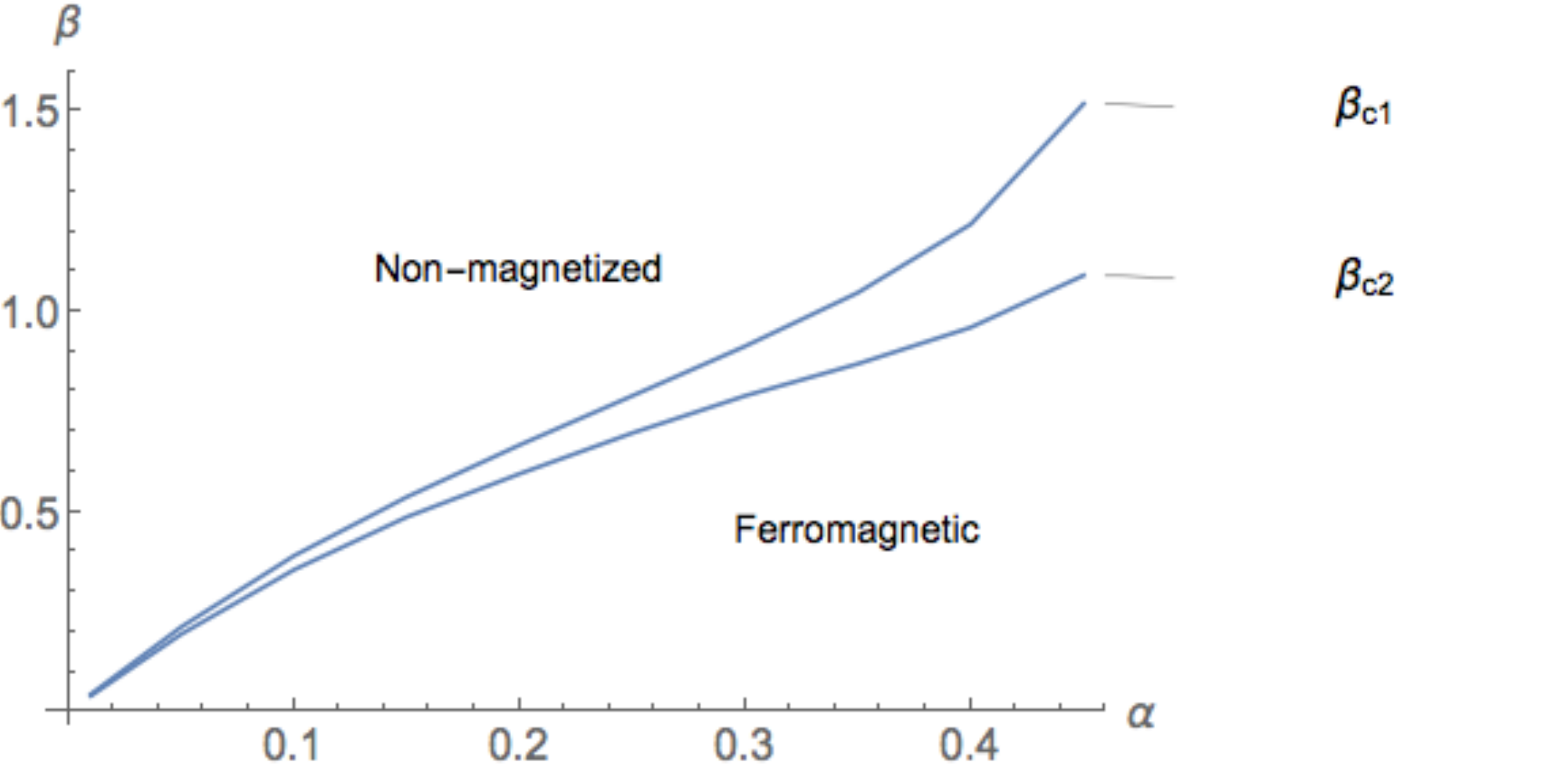}
  \caption{Phase plot in the fine structure constant $\alpha$ - $\beta=p_F/m$ plane 
showing the 
  allowed region of spontaneous ferromagnetism}
  \label{fig:phasedia}
\end{figure}

Thus we see that there are three possibilities, $\zeta=1$ is not a minimum at all, $\zeta=1$ is a local minimum but
$E(0)<E(1)$ i.e. a metastable vacuum and finally, $\zeta=1$ is the absolute minimum with $\zeta=0$ unstable vacuum.
In Fig.~\ref{fig:phasedia} we have plotted the approximate regions of the three phases in the parameter space.
An interesting fact about the above results is that the number density and the mass of the fermion 
determine the properties of the phase diagram only through the ratio $\beta\equiv p_F/m$.  
For the applicability of the formalism we need the fine structure constant to be in the 
perturbative range $\lesssim 0.2$. It can then be seen that $\beta $ is constrained to remain 
$\lesssim 0.5$. Thus the gas can be relativistic but not ultra-relativistic. Further, the 
formalism is valid only so long as the exchange energy is the dominant source of modification to 
the quantities for the free non-interacting gas. 
\begin{table}[tbh]
\begin{centering}
\begin{tabular}{|c|c|c|c|c|}
 \hline
Fine structure constant & $p_F/m$ & Energy density $E(\zeta=1)$&
$\Delta E= $E(0)-E(1)$ $& Rest mass energy density\\
 $\alpha_{X}$ &  & in $m^4$ units&  in $m^4$ units& in $m^4$ units\\
 \hline
 $0.01$ & $0.01$& $-1.618\times 10^{-9}$& $5.4\times 10^{-11}$ & $2.162\times 10^{-6}$ \\
 $0.05$&$0.02$& $-9.70\times 10^{-10}$&$1.90\times 10^{-10}$& $2.702 \times 10^{-7}$\\
 $0.10$&$0.10$ &$-1.12\times 10^{-6}$& $2.1\times 10^{-7}$& $3.38\times 10^{-5}$\\
 $0.10$&$0.30$ &$-5.84\times 10^{-5}$& $5.3\times 10^{-6}$& $9.12\times 10^{-4}$\\
 \hline
\end{tabular}
\end{centering}
\caption{
Representative values of energy density difference $\Delta E$ between the unfavourable spin 
balanced state with $\zeta=0$ and the ferromagnetic favoured state with $\zeta=1$.  It is a measure 
of the excess energy density stored in the domain walls. $m$ is the mass of the relevant fermion. 
The last column contains the rest mass energy density.}
\label{table:deltae}
\end{table}
In Table \ref{table:deltae} we list representative values of the parameters that are favourable for spontaneous
ferromagnetism. The rest mass energy density, the last term subtracted off in \eqref{eq:ekin} is also listed
for comparison. 

In understanding the cosmological behaviour of PAAI it is useful to understand the relative importance of the 
various contributions to the energy density. We expand the corresponding  expressions in the powers of $\beta$. 
There are three contributions. Kinetic energy density of the free gas, but without the rest mass, the exchange 
energy density and the rest mass energy density. We also see from our diagrams that the local minima occur 
only at $\zeta=0$ or at $\zeta=1$. So we
expand the expressions \eqref{eq:ekin} and \eqref{eq:exchange} for the two critical values 
of $\zeta$ for small $\beta$.
For $\zeta=0$, we have
\be
E^\text{kin}(0)
=m^4\left\lbrace\frac{\beta^5}{10 \pi ^2}-\frac{\beta^7}{56 \pi ^2}
+ O\left(\beta^{9}\right)  \right\rbrace
\ee

\be
E^\text{xc}(0)=
-\alpha_{X} m^4 \left\lbrace \frac{\beta^4}{\pi ^2}+\frac{2 \beta^6}{3 \pi ^2}
+O\left(\beta^{8}\right) \right\rbrace
\ee

For $\zeta=1$ we have 
\be
E^\text{kin}(1)
=m^4\left\lbrace \frac{\tilde{\beta}^5}{20 \pi ^2}-\frac{\tilde{\beta}^7}{112 \pi ^2}
+O\left(\beta^{9}\right) \right\rbrace
\ee
 
 \be
E^\text{xc}(1) =
-\alpha_{X} m^4 \left\lbrace \frac{{\tilde{\beta}}^4}{2 \pi ^2} - \frac{7 {\tilde{\beta}}^6}{27 \pi ^2}
+O\left(\tilde{\beta}^{8}\right)  \right\rbrace
\ee
where $\tilde{\beta}=2^{1/3}\beta$. Thus we see that the rest mass term dominates for small
$\beta$. While the kinetic energy and exchange energy contributions determine the favourable
collective state, the contribution to cosmologically relevant energy density is made by the rest 
mass term.

In the following sections we shall be considering PAAI medium in the context of cosmology where it may undergo
the ferromagnetic phase transition as the Universe cools from the Big Bang and later perhaps by the present epoch also
undergo a reverse transition to the symmetric phase. It is useful to know the effective equation of state of such a gas. 
Since the only local minima occur at $\zeta=0$ or at $\zeta=1$, it is sufficient to study these cases.  The energy density, 
designated $\rho$ in the cosmological setting,  can be obtained in these cases by setting the corresponding value 
of $\zeta$ in \eqref{eq:exchange}. 
The pressure is defined as and can be calculated from this through \cite{Chin:1977iz}
\be
p= n\frac{d {\rho}}{d n}-\rho \equiv \frac{1}{3}p_F\frac{d\rho}{dp_F} -\rho
\label{eq:pressuredefinition}
\ee
whose expression is not explicitly displayed here.
\begin{figure}[tbh]
  \includegraphics[width=\linewidth]{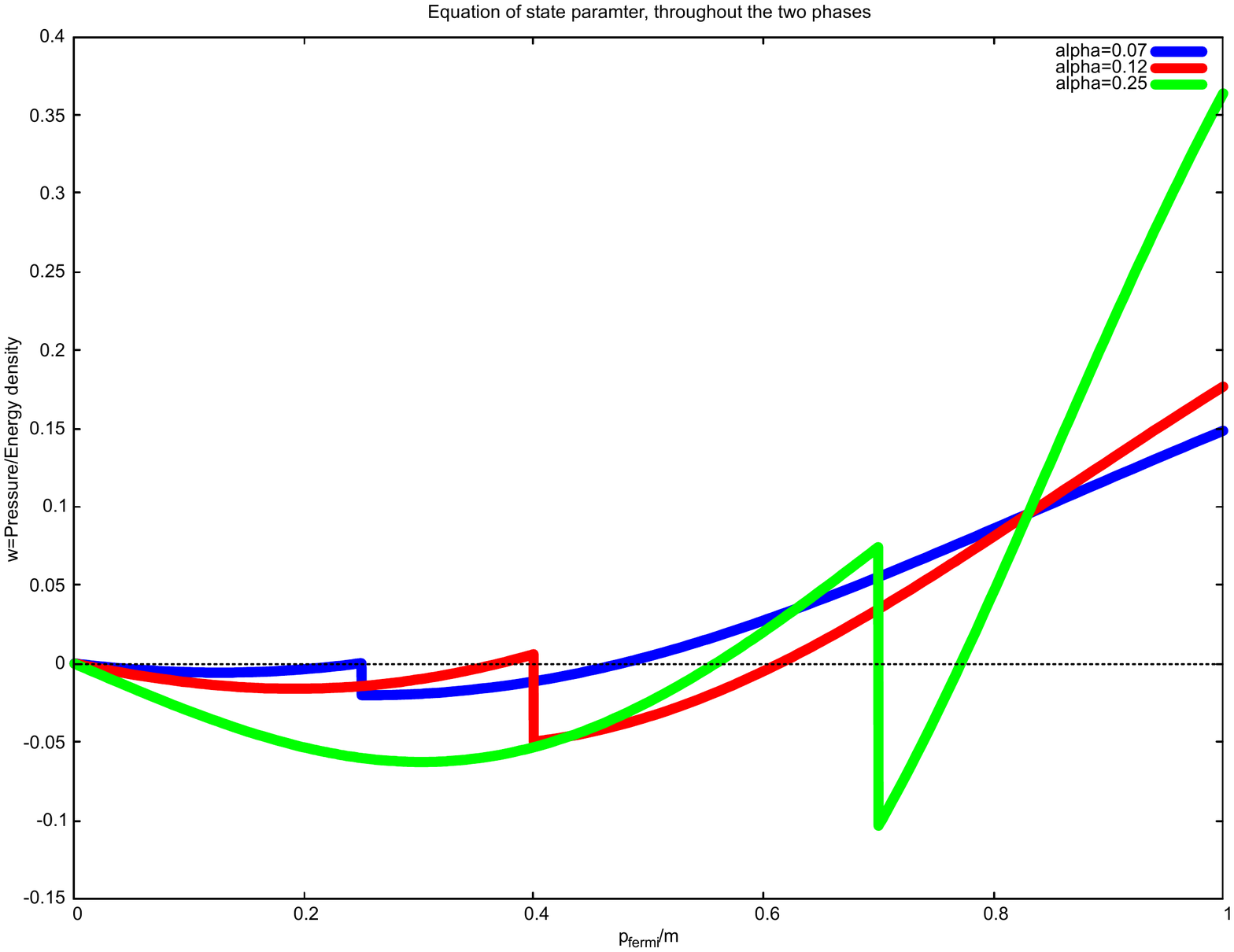}
  \caption{Variation of the effective equation of state parameter $w$ of the PAAI as a function of the parameter $\beta=p_F/m$.
  The medium in each case passes from the ferromagnetic phase at lower values of $\beta$ to non-magnetised
  or symmetric phase at higher values of $\beta$ as taken from Fig.~\ref{fig:phasedia} 
}
  \label{fig:eos-compre}
\end{figure}

In Fig.~\ref{fig:eos-compre} we plot the behaviour of the parameter $w=p/\rho$ for the PAAI.
We use representative values of the coupling $\alpha$, $0.07$, $0.12$ and $0.25$,  but not much larger by the need to 
remain within the perturbative validity of the calculation. 
The figures take account of the restrictions indicated by the phase diagram Fig.~\ref{fig:phasedia} on the range of 
$\beta=p_F/m$ parameter within which the medium is either in the ferromagnetic or the symmetric states. 
We see that over the range of $\beta$ values of interest, $w$ varies over $-0.1<w<0$. This mildly negative equation of
state parameter is qualitatively similar to the case of extended objects discussed in Sec.~\ref{sec:negpress} and
suggests a regime of strong correlations. We shall 
see that this parameter itself is not relevant to cosmology. What we learn is that the exchange energy plays a significant 
role in the behaviour of PAAI. We shall be assuming that this collective effect makes the medium immune to the much 
weaker effects of the cosmological gravitational field.

\section{Domain walls}
\label{sec:dw}
Macroscopic ferromagnetic systems are characterized by the occurrence
of domains. The possible directions of spontaneous magnetization are associated 
with preferred directions in the crystal.  We shall now parameterize the
the physics that governs cosmic ferromagnetic domains.
The domain walls are
not expected to be topologically stable. This is because the underlying
symmetry is $SU(2)$ of spin, which permits rotations within the vacuum
manifold for the defect to disentangle. However these processes are
suppressed by a competition between the gradient energy and the
extra energy stored in the domain walls as detailed below.  

The net magnetization in volume $V$ is given by $\vec{M}_V =\mu_M \int_V d^3x 
\langle \bar{\Psi}\vec{\Sigma} \Psi\rangle$ were $\mu_M$ is the intrinsic magnetic moment of the magnino and
$\Psi$ is the fermionic field operator 
and $\Sigma^i=\frac{1}{2} \epsilon^{ijk}\gamma^j \gamma^k$  in Dirac formalism.
Introduce the local vector order parameter $\vec{S} \sim \langle \bar{\Psi}
\vec{\Sigma} \Psi\rangle$
up to a dimensional constant to make $\vec{S}$ canonically normalized.  
The Landau-Ginzburg effective lagrangian for $\vec{S}$ is
\be
\label{eq:leffS}
\mathcal{L}\  =\ \frac{1}{2}\partial_\mu \vec{S}\cdot \partial^\mu \vec{S}
- \frac{\lambda}{8}(\vec{S} \cdot \vec{S}- \sigma_1^2)^2 + \ldots
\ee
where the  dots denote irrelevant terms.  The value $\sigma_1$ sets the scale of the
magnetization in the medium. Over uncorrelated distances, the magnetization
will settle to different orientations. 
The critical temperature of this phase transition is 
$T_c\sim \sigma_1$. The structure of these walls 
freezes in at  Ginzburg temperature \cite{Kibble:1980mv} 
$T_G$ given by $(T_c - T_G)/T_c=\lambda^2$.  Below this temperature there
isn't sufficient free energy available to disrupt the order within volumes of 
size $\xi_{G}^3$ where $\xi_G=1/(\lambda^2 \sigma_1)$.

A domain wall separating two such regions
can be thought of as a constrained soliton. The Lagrangian \eqref{eq:leffS}
signals a non-zero expectation value for the modulus of the order parameter. This gives
rise to  Goldstone bosons, the magnons. However the medium is not translationally
invariant due to occurrence of domain walls. The walls are  
a transition region in which the direction of the condensate is changing, ie,
the angular parts of the order parameter also acquire a vacuum expectation value.
The phenomenon, given the boundary condition stated above can be
self consistently described by introducing the field 
$\theta(\vec{x})=\arccos(\hat{S(\vec{x}}) \cdot \hat{S}_1)$ with respect to a fiducial $\hat{S}_1 $. 
This $\theta$  interpolates and acquires the value $\theta_{12}=\arccos 
\hat{S}_1 \cdot \hat{S}_2$ as we transit from one domain to the next. Since the energy 
density due to misaligned spins is proportional to $\cos \theta$,
this results in  a sine-Gordon type lagrangian for $\theta$, 
\be
\label{eq:leff}
\mathcal{L} = \frac{1}{2}\sigma_1^2 
\partial_\mu\theta \partial^\mu \theta
- \frac{\sigma_2^4}{\kappa^2}(1 - \cos \kappa\theta) + \ldots  
\ee
where $\kappa= 2\pi/|\theta_{12}|$ fixes the period  and $\sigma_2$ is another
parameter of the dimension of mass. Such domain walls have width
$w=\sigma_1/\sigma_2^2$ and energy per unit area $\mathcal{E}/A=8\sigma_1\sigma_2^2/\kappa^2$.

These consideration permit in principle matching of the mean field theory for the domain walls with
the gross parameters needed in cosmology. We shall be working with the network  characterised by the parameter $\omega$ 
which is the width of individual domain walls and with average separation between walls given by a parameter $L$. 
As a source in Friedmann equation we shall be  using the energy of domain walls averaged over a
large number of domains.

\subsection{Evolution and stability of domain walls}
\label{sec:dwstability}
One of the main sources of wall depletion is mutual collisions.
However the walls become non-relativistic below the Ginzburg  
temperature and the free energy available for bulk motion reduces.
A further source of instability is 
from the spontaneous nucleation of the trivial vacuum regions, in the
form of holes in the wall, bounded by a string-like
defect. This mechanism has been studied in detail in \cite{Preskill:1992ck}.
The rate for such decay is governed by an exponential factor $\exp(-B/\lambda)$
\cite{Kobzarev:1974cp} where the exponent is the Euclidean action of the "bounce" solution
connecting the false and the true vacua \cite{Coleman:1977py}. The part $B$ is
oder unity and if the coupling $\lambda$ as introduced in the Landau-Ginzburg
action \eqref{eq:leffS} is small the requisite stability is possible. 
On phenomenological grounds we need this complex to be stable for
several billion years, or $\approx 10^{17}sec$. Since large suppression factors $\sim10^{-30}$
$\approx e^{-69}$ are natural for $\lambda \sim 0.01$, we can assume the intrinsic stability of such walls over 
the required epochs. The realistic mechanism for disintegration of the DW network resides in the magnino gas 
becoming non-degenerate.

\section{A minimal model for Dark Energy}
\label{sec:singlemagnino}
We consider a hitherto unobserved sector with particle species we generically call $M$ and
$Y$. They are assumed to be charged under a local abelian group $U(1)_{X}$ with fine structure constant $\alpha_{X}$. Their charges $Q_{X}$ in the simplest case are equal and opposite, $Q_{X}(M)=-Q_{X}(Y)$.  The two are also assumed to be distinguished by
an additional charge $G$ which is global, or could be a parity, such that mutual annihilation is
forbidden, in a way analogous to the role of the charge $B-L$ in the observed sector.
 The species $M$ is assumed to have very small mass $m_M$ in the sub-eV range  and is  referred to as \textit{magnino} 
 because as we shall explain, the magnetised collective state is its hallmark property.
The $Y$ mass $m_Y$ is assumed to be much larger. With charges as assigned here, neutrality requires that
the number densities of the two species have to be equal, in turn this means that the Fermi energies
are also the same. The hypothesis of larger mass is to ensures that $Y$ does not enter into a collective magnetic phase.

Furthermore, assuming a parallel Big Bang history, there are the $U(1)_{X}$ photons at a temperature $T'$ in this sector.  
We need to place desirable requirements on the values of this temperature at the current time and over the time period for which DE is important. The requirements are as follows. 
Recall that $\beta\equiv p_F/m_M$.
\begin{description}
\item[T1] The need for the $M$-$Y$ system to remain a plasma, so that $T'$ is larger than the Hydrogen like 
binding energy of the system. 
\be
T'> \alpha_{X}^2 \frac{m_M m_Y}{m_M + m_Y} \approx \alpha_{X}^2 m_m 
\ee
\item[T2] The gas of the magnino particles $M$ is quasi-relativistic, i.e. $\beta\approx 1$, and degenerate so that 
\be
T' \ll E_F^M \implies \frac{T'}{m_M} \ll \sqrt{\beta^2 +1} - 1
\ee
\item[T3] The gas of the $Y$ particles is non-relativistic and non-degenerate. Thus we assume $E_F^Y\approx {p^2_F}/{2m_Y}$,
and 
\be
T' \gg E^Y_F \approx \frac{1}{2}\left( \frac{m_M}{m_Y} \right) m_M \beta^2
\ee
\end{description}
Requirement \textbf{T3} is natural if the ratio of the masses $m_M/m_Y $remains small. 
In a detailed analysis these requirements would help narrow down the actual parameters of this sector.

In using the quantities we considered in Sec.~\ref{sec:rfg}, we will need to identify
\be
\rho^M=E(\alpha_{X},\beta,\zeta) + m_M^4\frac{\beta^3}{(3\pi^2)}
\label{eq:rhocosmX}
\ee
where we reserve $\beta=\pF/m_M$ to stand for this parameter as referring to $M$ particles, 
the parameter $\zeta$ will take values $0$ or $1$, and we have added back the rest energy,
the last term subtracted in \eqref{eq:ekin},  since now we are considering the
response to gravity. As discussed at the end of Sec.~\ref{sec:phasediaeos} the rest mass term dominates for small $\beta$. While the kinetic energy and exchange energy contributions determine the favourable collective state, the contribution to cosmologically relevant energy density is made by the rest mass term.

We  start our considerations at time $t_1$  when the temperature is just below
$T_G$ so that the wall complex has materialised. In the following we are going to ignore temperature effects, 
which is a valid assumption when the conditions \textbf{T1-T3} above are satisfied, however if needed at other
epochs where these conditions change, it is easy to extrapolate within the Big Bang paradigm from conditions 
already established.  The parameters of this wall complex are $\omega$, the thickness of individual 
walls and $L$ the average separation between walls, as introduced in Sec.~\ref{sec:dw}. 
Under these circumstances, our first observation is that the wall complex is stabilised by the  magnetic forces, 
and being much smaller than the scale  of the horizon, is unaffected by the cosmic expansion. Then on the scale of
the horizon, the wall complex behaves just like a space filling homogeneous substance and the situation is 
that pictured in Fig.~\ref{fig:scalingvacuum}.  Further, due to the demand of neutrality, the heavier gas $Y$ cannot
expand either, although it has no condensation effects. Let us denote the number density of the magninos
trapped in the walls to be $n^X_{\mathrm{walls}}$ and the remainder residing in the enclosed domains
by $n^X_{\mathrm{bulk}}$. Averaged (coarse grained) over a volume much larger than the $L^3$, this gives the
average number density of the magninos to be
\be
\langle n^X \rangle = \frac{\omega}{L} n^X_{\mathrm{walls}} + \left( 1- \frac{\omega}{L} \right) n^X_{\mathrm{bulk}}
\ee 
And from the neutrality condition we have
\be
\langle n^X \rangle = \langle n^Y \rangle
\ee
Then we can demand that PAAI in this phase acts as the DE, so that assuming $Y$ to be non-relativistic, 
and ignoring other contributions,
\be
\rho^Y \approx m_Y \langle n^Y \rangle = \rho_{\text{DE}} = 2.81\times10^{-11} (\eV)^4
\ee
Expressing the number density of $Y$ as a ratio of the number density $n_\gamma=3.12\times 10^{-12}(\eV)^3$ 
of photons, we set $\eta^Y= \langle n^Y \rangle/n_\gamma$. Then we obtain 
\be
m_Y \approx \frac{1}{\eta^Y} 12.9 \eV
\ee
Then the Fermi momentum of  both $M$ and $Y$ is
\be
p_F= (3 \pi^2 \langle n^Y \rangle)^{1/3} = (\eta^Y)^{1/3}4.52\times 10^{-4} \eV
\ee
Then 
\be
\beta^Y\equiv \frac{p_F}{m_Y} =  (\eta^Y)^{4/3} 3.50\times 10^{-5} 
\ee
and for the magnino we obtain
\be
\label{eq:SIbeta}
\beta = (\eta^Y)^{1/3}  \left( \frac{\eV}{m_M}\right) 4.52\times 10^{-4} 
\ee
We can now make a wish list of propositions characterising the first of our two scenarios, with prefix SI. 
(Later we introduce another scenario SII).
\begin{description}
\item[SI-1] The magnino species $M$ of mass $m_M$ is a degenerate gas and 
undergoes ferromagnetic condensation. Its density does not scale with the expanding universe.  
\item[SI-2] The species $Y$ has mass $m_Y\gg m_M$ and does not undergo condensation.
\item[SI-3] The $Y$ gas remains tied to the magnino condensate for neutrality and its density also does not scale. 
Thus $M$ and $Y$ together, but dominated by the $Y$ mass,  simulate the DE.
\item[SI-3] The fine structure constant of this hidden  sector has a value ensuring validity of perturbation theory,
$\alpha_{X}\lesssim 0.2$. In order to obtain ferromagnetic condensation as in \textbf{SI-1}, we must have $\beta \lesssim 0.5$
as per phase diagram Fig.~\ref{fig:phasedia}. For degeneracy we expect $\beta$ to be $O(1)$. 
We will assume that $\beta$ is largest allowed by the phase diagram for a given $\alpha_{X}$.
\end{description}

Then to ensure \textbf{SI-3}, using \eqref{eq:SIbeta}, we have
\be
\label{eq:SImlowerbound}
m_M \gtrsim (\eta^Y)^{1/3}\left(\frac{0.1}{\beta} \right)4.52 \times 10^{-3}  
\ee
where $\beta$ will not change appreciably over the range of interest of $\alpha_{X}$.
These conditions together determine the ratio
\be
\label{eq:SImupperbound}
\frac{m_M}{m_Y}=\frac{\beta^Y}{\beta}\approx (\eta^Y)^{4/3} \times 10^{-6} \ll 1
\ee
\begin{figure}[tbh]
\includegraphics[height=0.75\textheight]{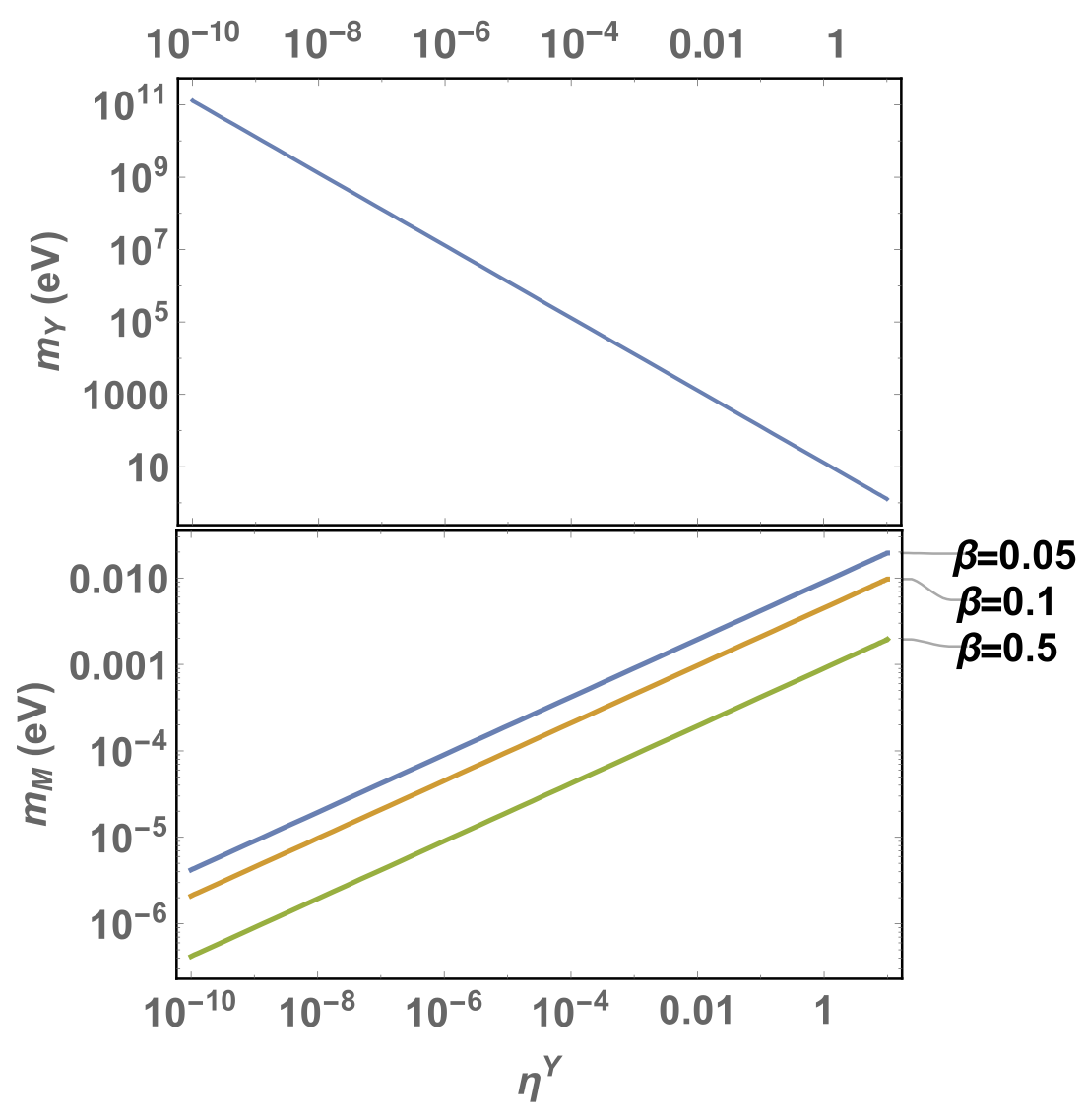}
  \caption{Allowed masses $m_Y$ and $m_M$ as a function of $\eta^Y$. }
  \label{fig:masses_vs_etaY}
\end{figure}
and the last inequality is meant to impose the requirement \textbf{SI-2}.
If we implement \textbf{SI-2} by demanding that $\beta^Y\lesssim 10^{-3}$, then $\eta^Y\lesssim 10^{-2}$.
In Figure \ref{fig:masses_vs_etaY}  we show the region allowed by above considerations in the $m_M$-$\eta^Y$ 
and $m_Y$-$\eta^Y$ planes.

\section{A flavoured model with a solution to the concordance puzzle}
\label{sec:flavo}
It is now interesting to explore whether this hidden sector admitting $X$-ferromagnetic condensation mechanism 
also has a candidate for the  DM. In scenario \textbf{SI}, although two mass scales exist, one of them only ensures
degeneracy and condensation, while the other mass scale gets tied to the DE scale. Also, most crucially,
these species do not dilute with the expansion of the universe in order to mimic constancy of vacuum energy. In order 
to explain DM we need additional species whose abundance can scale and dilute like non-relativistic matter.
 
Let there exist several stable species $M_a$ and $Y_a$ with $a$ the "flavour" index. Let their abundances be
denoted, with a simplification in notation as $\eta^{M1}$, $\eta^{Y1}$  ... $\eta^{Ma}$, $\eta^{Ya}$ etc.
Let us denote the \textsl{general} requirements to be obeyed by such \textsl{flavoured} scenarios to be \textbf{GF}.
The wish list of such requirements is

\begin{description}
\item[GF1] The charges of these species under $U(1)_{X}$ are opposite in sign for $M$-type versus $Y$-type. 
However we leave open the possibility that the magnitudes of these charges can be small integer multiples of each other.
\item[GF2] The  heavier species of $M$-types and $Y$-types should be stable against decay into the corresponding 
lighter ones even if their $Q_{X}$ charges tally. This is analogous to flavour symmetry in the observed 
sector, where the purely electromagnetic  conversion of heavier leptonic  flavours into lighter ones is not observed.   
\item[GF3] The lightest pair $M_1$ and $Y_1$ (more generally at least one effective degree of freedom of species of each type) 
have  equal and opposite charges, and satisfy the requirement of scenario \textbf{SI} so that DE is accounted for. 
\item[GF4] The heavier species (more generally the remainder degrees of freedom) do not undergo condensation.
\end{description}

Within these general criteria the simplest scenario that can be thought of may be called \textbf{FI}. It has the 
following straightforward requirements 
\begin{description}
\item[FI-1] The pair of species $M_2$ and $Y_2$ with $Q_{X}(M_2)=-Q_{X}(Y_2)$
\item[FI-2] This pair of species accounts for the observed DM.
\end{description}
Thus we demand, with $n_{M2}=n_{Y2}$ designating the number densities, that
\be
(m_{M2}+m_{Y2})n_{Y2} = \rho_\text{DM}= 1.04\times 10^{-11} (\eV)^4
\ee
so that  
\be
m_{M2}+m_{Y2}=\left( \frac{1}{\eta^{Y2}}\right) 3.33 \eV
\label{eq:DMconstraint}
\ee
In order for either of  $X_2$ or $Y_2$, or both together to act as DM, the right hand
side of the above equation has to be at least a few keV (see Sec.~\ref{sec:cosmosummary}).
Thus we need 
\be
\eta^{Y2}\lesssim 10^{-3} \qquad \text{to ensure DM mass}\gtrsim \text{keV}
\label{eq:etaYconstraint}
\ee
From Sec.~\ref{sec:singlemagnino}, we have that $\eta^Y$ can take on any value $\lesssim 10^3$ and account for
DE adequately. The DM constraint on the second flavour restricts its abundance to $\lesssim 10^{-3}$.
In this scenario $\eta^Y$ and $\eta^{Y2}$ need not be related, and a few orders of magnitude difference in abundance 
could be easily explained by dynamics occurring within that sector in an expanding universe.
Further we shall see later that the large value of $\eta^Y$ makes the scenario capable of explaining 
the origins of cosmic magnetic fields, while the small $\eta^{Y2}$ value can separately solve the DM puzzle. 

The scenario $FI$ requires that at least one of $M_2$ and $Y_2$ is heavy enough to be the DM particle.
But it leaves the mass of the other particle undetermined. A scenario that is more restrictive about the mass of $\eta^{M2}$ could arise 
as follows, and we denote this scenario \textbf{FII}. 
\begin{description}
\item[FII-1] There are two species $M_1$ and $M_2$, of the same charge $Q_{X}(M_2)=Q_{X}(M_1)$.
\item[FII-2] $\eta^{M1}= \sigma \eta^{M2}$  where $\sigma$ is a numerical factor
\item[FII-3] Only $M_1$ is the magnino, capable of condensing.
\item[FII-4] There is only one species of $Y$ type, with $Q_{X}(Y)=-Q_{X}(M_1)$.
\end{description}
For neutrality of the medium we need $\eta^Y=\eta^{M1}+\eta^{M2}$. Then in this scenario, the fraction equivalent to
$\eta^{M1}$  of the $Y$ particles will suffice to keep the condensed state of $M_1$ neutral, and thus the mass of $Y$ will 
be determined as in \textbf{SI}. The remainder $Y$ particles, in abundance $\eta^M2$ scale like free matter particles.
Then analogous to conditions Eq.s \eqref{eq:DMconstraint} \eqref{eq:etaYconstraint}, we get
\bea
m_{M2}+m_{Y}&=&\left( \frac{1}{\eta^{Y2}}\right) 3.33 \eV \\
\eta^{M2}&\lesssim& 10^{-3}
\eea
The point is that $m_{Y}$ is already determined by the value of $\eta^{M1}$ from DE Condition, and if 
$\eta^{M1}\gtrsim 1$  then mass of $Y$ would be determiend to be too small to be DM candidate.
In this case, without proliferating unknown mass values, $m_{M2}$ can be the DM candidate.

This Dark Matter sector is along the lines of \cite{2009-Feng.etal-JCAP}, and through out its history could have been
partially ionised and could be progressively becoming neutral. In particular it represents the class of self interacting
Dark Matter including van der Waals forces that may result between such atoms due to very low binding energy.  
It has  been argued for example in  \cite{2017-Kamada.etal-Phys.Rev.Lett.} that such a model potentially explains
the diversity in the rotation curves of galaxies.

\section{Origin of cosmic magnetic fields}
\label{sec:cosmagfields}
The origin and evolution of galactic scale magnetic fields is an open
question \cite{Kulsrud:1999bg,Kulsrud:2007an}. In particular the extent of seed magnetic 
field as against that generated by subsequent motion is probably 
experimentally distinguishable \cite{Durrer:2013pga}\cite{Subramanian:2015lua}. 
In the present case, we can estimate the field strength of the $X$-magnetism in each domain.
Using the leading term from \eqref{eq:magnetisation} and using the value $\mu_0 \mu_B=1.5 \times 10^{-8} T(\eV)^{-3}$,
\bea
B_\text{dom}=\mu_0 M_\text{dom}\ &\approx& 
\mu_0\left( \frac{m_e}{m_M}\right)\left( \frac{e'}{e}\right)\mu_B \left( \frac{m_M^3}{6\pi^2}\right)
\times \frac{1}{6}\times(2^{1/3}\beta)^3 \\ \nonumber
&\approx&  \left(\frac{m_M}{\eV}\right)^2 \left( \frac{e'}{e}\right)\left(\frac{\beta}{0.1}\right)^3 \times 2.2\times10^{-8} T
\eea
Since the domain structure is completely random we expect zero large scale magnetic field on the average. 
Residual departure from this average can be estimated by assuming that the deviation from the mean grows as $\sqrt{N}$ as
we include $N$ domains. Thus if the $X$-magnetic field in individual domains has the value $B_\text{dom}$ then on the scale of 
galactic clusters  $L_\text{gal}$ it possesses a root mean square value  $\overline{\Delta B}\equiv B_\text{dom}(L/L_\text{gal})^{3/2}$. 
\begin{figure}[bht]
  \includegraphics[width=\linewidth]{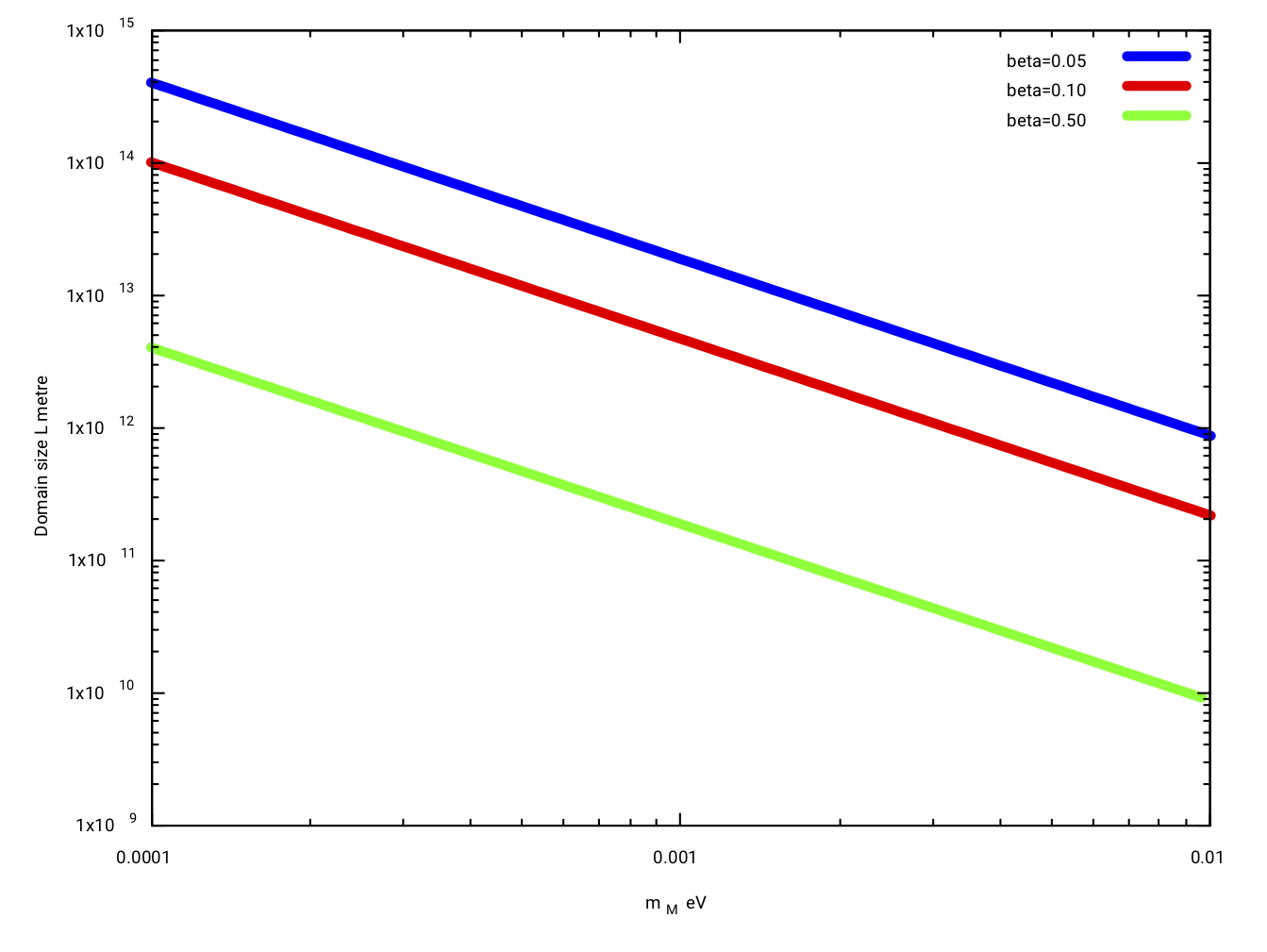}
  \caption{Desirable average size $L$ of the $X$-ferromagnetic domains as a function of magnino mass $m_M$ so as 
  to provide required seed for observed intergalactic magnetic fields. The parameter beta stands for $p_F/m_M$ as before.
  The ratio of hidden charge to standard charge $e'/e=1$ is assumed for illustration.}
  \label{fig:domainsizevsmM}
\end{figure}

Assuming $U(1)_{X}$ field mixes kinetically with standard electromagnetism through term of the form 
$\xi F^{\mu \nu}F^X_{\mu \nu}$, the $\xi$ is well constrained from Supernova 1987A data to\cite{2000-Davidson.etal-JHEP}
$10^{-7}<\xi < 10^{-9}$. On the other hand from the CMB data the relative energy density contribution $\rho_{\mathrm{mcp}}/\rho_{cr}$
to the cosmic budget is shown to be constrained to \cite{2013-Dolgov.etal-Phys.Rev.} $\equiv\Omega_{\mathrm{mcp}}h^2<0.001$.
More recently \cite{2018-Chang.etal-JHEP} reports exclusion of all mass values $<200$MeV based on the supernova
1987A data. We proceed here to make an estimate based on our model, compatible with all constraints except that
of the last mentioned paper, pending further verification of that constraint.
The exact value of the seed required depends on the epoch at which they are being studied and
other model dependent factors\cite{2012-Widrow.etal-SpaceSci.Rev.}.  We consider the the possibility of a 
seed of $10^{-30}$T with a coherence length of $0.1$ kpc$\sim3\times 10^{18}$ metre obtained with $\xi=10^{-8}$.
\be
\overline{\Delta B}_{\mathrm{seed}}=10^{-30}T\sim 10^{-8}\times\left(\frac{m_M}{\eV}\right)^2\left( \frac{e'}{e}\right) \beta^3 
\left( \frac{L}{\mathrm{metre}}  \right)^{3/2}\times 10^{-40} T 
\ee
In Fig. \ref{fig:domainsizevsmM} we show the values of $m_M$ and $L$ that can potentially satisfy
this requirement, setting $e'/e=1$ for simplicity. It can be seen that representative values for $L$ for $\beta=0.1$ are 
in the range  $10^{11}$-$10^{13}$ metre which is solar system size. A detailed treatment to estimate the residual
fluxes on large coherence length scales could trace the  statistics of flux values in near neighbour domains and the rate at which 
the magnetic flux could undergo percolation, providing perhaps a smaller value for $L$.

\section{Conclusions}
\label{sec:conclusion}
Dark Energy problem is sufficiently important that it is worth exploring all avenues to its explanation. We have proposed
a specific mechanism from known many body physics that may be of relevance to understanding this enigmatic
phenomenon. The two known mass scales of elementary particle physics, $\Lambda_{QCD}$ and the Standard Model
Higgs vacuum expectation value do arise as effectively non-perturbative effects, but are clearly not at work since they are 
so much larger than the required energy scale.  On the other hand particle species of very light mass are now established
to exist. That makes it natural to inquire whether an alternative collective phenomenon involving suitably light particle 
species and new gauge forces which are not a part of standard particle physics, is at work. By assuming the existence 
of an autonomously lighter mass scale set by DE to be arising from masses of the new particles, we avoid having to 
explain it, at least at this stage of development.  On the other 
hand, the peculiar equation of state  can be deduced as an outcome of nothing more radical than an unbroken abelian 
gauge force.  Extended and space filling objects, specifically domain walls as possible solutions to understanding 
Dark Energy have  been proposed earlier in a variety of scenarios  as well \cite{Battye:1999eq,Battye:2007aa}\cite{Conversi:2004pi}\cite{Friedland:2002qs} \cite{Yajnik:2014eqa}.

Ferromagnetic state is a strongly correlated one, but can be understood within the fermi liquid framework and affords connecting
the collective observables to the microscopic constants and parameters. That it can occur in the presence of periodic 
translationally symmetric lattice has been known for long \cite{IbLu} as \textit{band ferromagnetism} or \textit{itinerant electron} phenomenon. However extending it
to the completely homogeneous situation, in fully relativistic setting  has not been previously carried out. We have adopted the formalism
of \cite{Bu:1984} to deduce the existence of such a state at varying mass values $m_M$ for the hypothetical magnino and 
the gauge coupling $e'$ of the new $U(1)_X$. While the magnino undergoes condensation, the effective Dark Energy 
density is determined by the mass of the heavier, non-ferromagnetic parter $Y$ needed to keep the medium 
neutral.  An appealing by product of this phenomenon is the possibility of explaining the origin of the cosmic magnetic 
fields, as discussed in Sec. \ref{sec:cosmagfields}.

The possibility of a hidden broken and unbroken $U(1)_X$ has been extensively explored, specifically that its connection with
the standard electromagnetism may be manifested in the existence of  minicharged particles. Recent experiments at DAMIC
\cite{Aguilar-Arevalo:2016zop}  have placed limits on dark photons, and the DM program of MiniBooNE \cite{Aguilar-Arevalo:2018wea} 
has, together with previous experiments,  reported null results in a variety of rather appealing models of Dark Matter 
( a comprehensive discussion is in \cite{ArkaniHamed:2008qn})  that use hidden electromagnetism and explain all 
the features of Dark Matter including their origin as thermal relics \cite{Pospelov:2007mp}. A multi component Dark 
sector has also been a recurrent theme of many works. Our model allows introduction of such additional species, with 
comparable masses and abundances as needed to explain the cosmic energy balance, however we have not attempted 
to relate it so far to other observables or as solution to unexplained  observations. We leave this to future work, indeed 
since the specific models explored and constrained by \cite{Aguilar-Arevalo:2018wea} do not rule out other possibilities. 
Although our core ingredients are fermions, in the DM models of Sec. \ref{sec:flavo}, the heavier 
second flavours are permitted to form neutral atoms and be bosonic DM as well.

A specific prediction of this model is the evolutionary nature of the $w$ parameter \cite{2005-Jassal.etal-Phys.Rev.,2005-Huterer.Cooray-Phys.Rev.,2004-Hannestad.Mortsell-JCAP,2005-Hall.etal-Phys.Rev.Lett.}, as also the eventual extinction, of the Dark Energy. 
The onset of ferromagnetic state would produce $w=-1$ for that medium, however depending upon the epoch at
which this happens DE component may not be dominant. However over the later epochs where it is making its presence
felt it may already be in disintegration due to percolation of the magnetic fluxes as also due to the possible end to the degenerate
phase of the magnino gas. We would thus expect a beginning with $w=-1$, progressing to $w=-2/3$ as appropriate
for domain walls when they are not densely packed within the horizon, and subsequent rapid decay towards $w=0$.
The ongoing studies in this direction will provide a validation or disproof of the model.

In an attempt to highlight the potential utility of the PAAI to cosmology, specifically to DE and to cosmic ferromagnetism, we have
been agnostic about the earlier history of this sector. Specifically it will be important to determine the fate of this medium at
a non-zero temperature, understand the nature of its phase transition, and gain a handle on the length scale $L$ in the spirit of
Kibble-Zurek mechanism \cite{Zurek:1985qw, Zurek:1996sj}. Some hints as to the temperature of this medium may be gained
if it has manifested itself in the excess cooling in CMB observed in the cosmic dawn \cite{2018-Bowman.etal-Nature}\cite{2018-Barkana-Nature}\cite{2018-Munoz.Loeb-Nature}. 
These issues need further investigation  towards verifying the extent of utility of the mechanism we have presented here. 

\section{ACKNOWLEDGEMENTS}
We thank NSERC, Canada for financial support and the Ministère des relations internationales et la 
francophonie of the Government of Québec for financing within the cadre of the Québec-Maharashtra 
exchange.  RBM and MBP also thank IIT Bombay for financial support and hospitality.


\bibliographystyle{apsrev}

\bibliography{refstoner,refcfmJan2011,refstoner-cosmonew}

\end{document}